\documentclass[11pt]{amsart}
\title{Vortex cusps}
\author{Volker Elling}
\usepackage{calc}
\usepackage{graphicx}
\usepackage{color}
\usepackage{amsmath}
\usepackage{amsxtra}
\usepackage{amssymb}

\renewcommand{\Re}{\operatorname{Re}}
\renewcommand{\Im}{\operatorname{Im}}
\newcommand{\csep}{\quad,\quad}
\newcommand{\eps}{\epsilon}

\newcommand{\defm}[1]{\emph{#1}}
\newcommand{\subeq}[2]{\mathord{\underbrace{\mathop{#1}}_{#2}}}

\newcommand{\eqv}{\Leftrightarrow}
\newcommand{\impl}{\Rightarrow}
\newcommand{\graph}{\operatorname{graph}}

\newcommand{\vv}{\vec v}
\newcommand{\half}{\frac12}
\newcommand{\pvint}{\text{p.v.}\int}

\newcommand{\boi}[2]{{]#1,#2[}}

\newcommand{\topref}[2]{\overset{\text{\eqref{#1}}}{#2}}

\newcommand{\dotp}{\cdot}
\newcommand{\ndiv}{\nabla\dotp}

\newcommand{\upconv}{\nearrow}
\newcommand{\dnconv}{\searrow}
\newcommand{\conv}{\rightarrow}

\newcommand{\const}{\text{const}}
\newcommand{\eqs}{\sim}
\newcommand{\pd}[1]{\partial_{#1}}
\newcommand{\qiq}{\quad\impl\quad}

\newcommand{\phiinf}{\phi_\infty}

\newcommand{\tz}{\ttac z}
\newcommand{\tgam}{\ttac\gam}
\newcommand{\twv}{\ttac\wv}
\newcommand{\tvx}{\ttac v^x}
\newcommand{\tvy}{\ttac v^y}
\newcommand{\tx}{\ttac x}
\newcommand{\ty}{\ttac y}
\newcommand{\cxd}{\xi}
\newcommand{\yex}{\alpha}
\newcommand{\gex}{\beta}
\newcommand{\cxu}{\cx_*}
\renewcommand{\vec}[1]{\mathbf{#1}}
\newcommand{\vn}{\vec n}
\newcommand{\wve}{e}
\newcommand{\wez}{e_0}
\newcommand{\wec}{e_1}
\newcommand{\mv}{\mu}
\newcommand{\Nv}{N}
\newcommand{\xx}{\vec x}
\newcommand{\vort}{\omega}
\newcommand{\ttac}[1]{#1}
\newcommand{\ssac}[1]{#1}
\newcommand{\cxx}{{\ssac\xx}}
\newcommand{\cx}{{\ssac x}}
\newcommand{\cy}{{\ssac y}}
\newcommand{\cz}{\ssac z}
\newcommand{\gam}{\Gamma}
\newcommand{\cgam}{{\ssac\gam}}
\newcommand{\cw}{\cwv}
\newcommand{\cq}{\ssac q}
\newcommand{\cvq}{\vec\cq}
\newcommand{\cqy}{\ssac\cq^{\cy}}
\newcommand{\cqx}{\ssac\cq^{\cx}}
\newcommand{\cvv}{\ssac\vv}
\newcommand{\cvy}{\ssac v^{\cy}}
\newcommand{\cvx}{\ssac v^{\cx}}
\newcommand{\wv}{w}
\newcommand{\cwv}{\ssac\wv}
\newcommand{\cva}{\overline\cvv}
\newcommand{\cvs}{\cvv}
\newcommand{\cvi}{\underline\cvv}
\newcommand{\qxi}{\underline\cq^\cx}
\newcommand{\vxs}{\cvx}
\newcommand{\vys}{\cvy}
\newcommand{\vxi}{\underline\cvx}
\newcommand{\vyi}{\underline\cvy}

\newcommand{\vya}{\overline\cvy}

\begin{document}

\maketitle

\begin{abstract}
  We consider pairs of self-similar 2d vortex sheets forming cusps, equivalently single sheets merging into slip condition walls,
  as in classical Mach reflection at wedges.
  We derive from the Birkhoff-Rott equation a reduced model yielding formulas for cusp exponents and other quantities 
  as functions of similarity exponent and strain coefficient. 
  Comparison to numerics shows that piecewise quadratic and higher approximation of vortex sheets agree with each other and with the model. 
  In contrast piecewise linear schemes produce spurious results and violate conservation of mass, a problem that may have been undetected in prior work
  for other vortical flows where even point vortices were sufficient.
  We find that vortex cusps only exist if the similarity exponent is sufficiently large and
  if the circulation on the sheet is counterclockwise (for a sheet above the wall with cusp opening to the right), 
  unless a sufficiently positive strain coefficient compensates.
  Whenever a cusp cannot exist a spiral-ends jet forms instead; we find many jets are so narrow that they appear as false cusps.
\end{abstract}


\begin{figure}
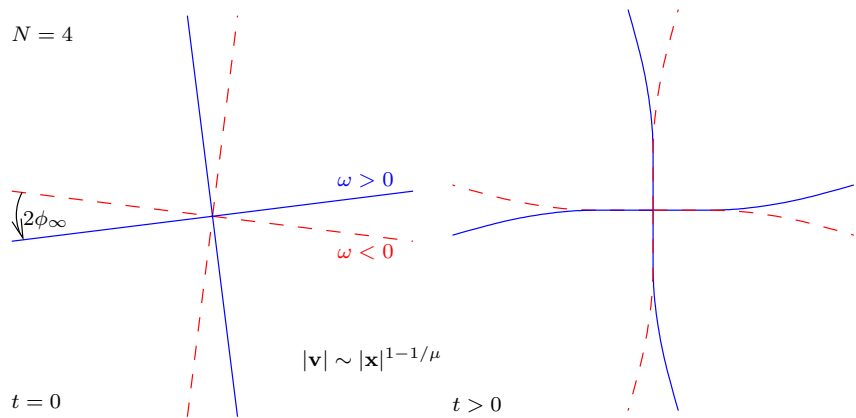

  \hfil\input{starini.pstex_t}\hfil\input{star.pstex_t}\hfil%
  \caption{Left: vortex sheet pairs; right: each pair forms a cusp}%
  \label{fig:Nsymm}%
\end{figure}

\section{Introduction}

This article studies planar self-similar incompressible vortex sheets meeting in cusps 
(fig.\ \ref{fig:Nsymm} right, fig.\ \ref{fig:cusp}). 
\defm{Self-similar} flows, also called ``pseudo-steady'' or ``quasi-steady'', have vorticity 
\[ \vort(t,\xx) = t^{-1} \vort(1,t^{-\mv}\xx)  \]
for some \defm{similarity exponent} $\mv$. The flow is ``similar'' at all times $t>0$, 
with any spatial distance $L$ dilated by a factor $t^\mv$ and any velocity $\vv$ scaled accordingly. 
Such flows arise from initial data of type
\[ \vort(0,\xx) = r^{-\kappa}\mathring{\vort}(\theta) \] 
where $r,\theta$ are polar coordinates. Since vorticity has dimensions of inverse time, the exponent $-\kappa$ of the initial singularity 
fixes $t^{-1}\sim L^{-\kappa}$ and thus determines the similarity exponent $\mv=1/\kappa$. 
To ensure velocity is locally integrable at the origin we only consider $\mv>\half$. 
The case $\mv=1$ is commonly studied for compressible flow as well since it corresponds to 
\[ \vv(t,\xx) = \vv(t^{-1}\xx). \] 

\begin{figure}%
  \includegraphics[width=.31\linewidth]{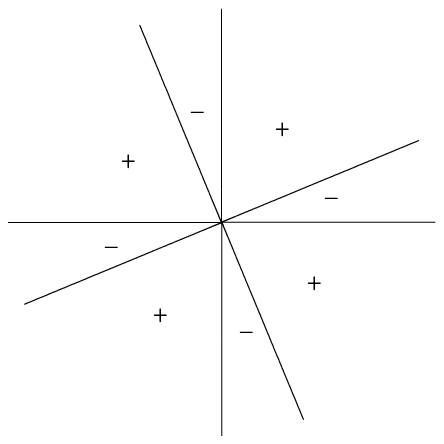}%
  \includegraphics[width=.31\linewidth]{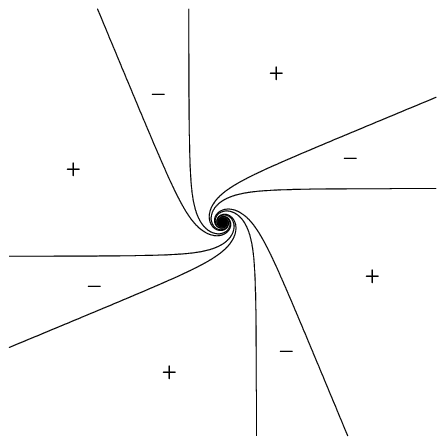}%
  \includegraphics[width=.31\linewidth]{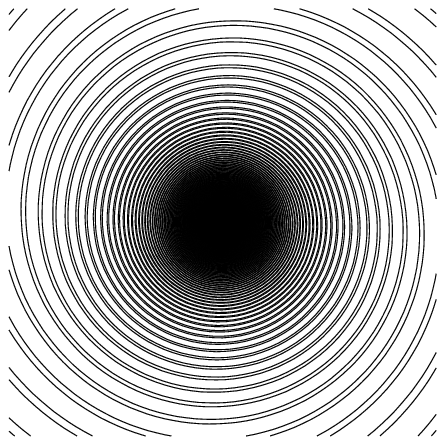}%
  \caption{Initial data (left) with mixed-sign $\vort$, flow for $t>0$ (center), $t\gg 0$ (right)}
  \label{fig:spirals}
\end{figure}

If $\vort$ is single-signed, or if the other sign is present but small enough to be ``dominated'', 
then strong rotation at the origin winds $\vort$ (in particular its $\vort=0$ level sets) into algebraic spirals of type $r\sim\theta^{-\mv}$ (fig.\ \ref{fig:spirals} center). 
For vortex \emph{sheets} algebraic spiral flows have been modelled in
\cite{kaden-1931}, \cite{rott-1956}, \cite{stern-1956}, \cite{birkhoff-1962}, \cite{mangler-weber}, \cite{moore-1975}; recently \cite{elling-mixedsignvort} gave a mathematically rigorous existence proof for a class of mixed-sign smooth $\mathring{\vort}$. 
(For logarithmic spirals with entirely different scaling see e.g.\ \cite{prandtl-1922}, \cite{alexander-vortex}, \cite{kaneda-doublespiral}, \cite{elling-gnann}.) 

It is natural to ask what happens in the borderline case where neither sign of vorticity dominates. 
In this article a pair of vortex sheets of equal strength but opposite sign is considered. 
It is shown that in some cases they may form a spiral-free cusp of type $y\sim\cxd^\yex$ (with $y$ distance to symmetry axis, $\cxd$ distance to cusp along axis);
otherwise a spiral-ends jet appears. 

\begin{figure}
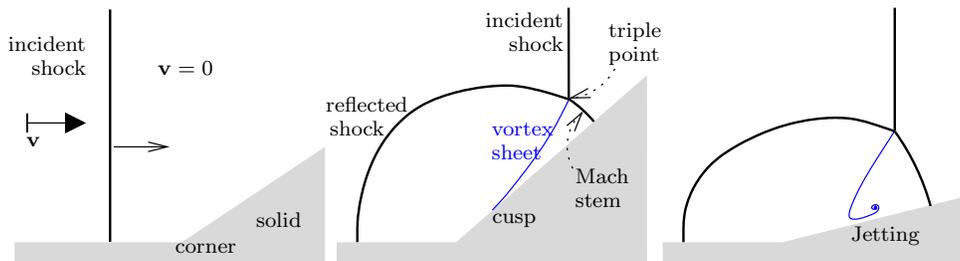

  \hfil\input{incident.pstex_t}\input{nojet.pstex_t}\input{jetting.pstex_t}\hfil%
  \caption{Self-similar simple Mach reflection favors strong jetting (right) for shallow ramps and large incident shock Mach numbers, 
    otherwise the sheet appears to blend into the wall in a cusp (center).}
  \label{fig:mr}
\end{figure}

While ``equal strength'' may at first glance appear to be a less important borderline case, 
it is precisely the case of a self-similar vortex sheet merging into a slip-condition wall (fig.\ \ref{fig:mr} center), 
since eliminating the wall by a reflection yields a mirror sheet of opposite vorticity.
Vortex-wall merging is commonly observed, for example in Mach reflection 
(fig.\ \ref{fig:mr}; see \cite{mach-wosyka}, \cite{neumann-1943}, \cite{hornung-reviews}, \cite{ben-dor-book}; experimental observations e.g.\ \cite{van-dyke} fig.\ 236, 237). 
\cite{henderson-vasilev-bendor-elperin-2003} and \cite{vasilev-bendor-elperin-henderson-2004} have begun a theoretical and numerical study on when the sheet merges smoothly rather than flipping backwards and 
forming a spiral-end jet near the wall.

In section \ref{section:modelling} we discuss several heuristic ODE models derived from the Birkhoff-Rott equation. 
Section \ref{section:vx} approximates axis-parallel on-sheet velocity $\cvx$ as a function of $\cxd$ alone, deriving a formula $\cvx\eqs\cxd^{\gex-1}$ (see \eqref{eq:gex})
with exponent $\gex$ depending only on $\mv$ and strain parameter $\wec$, a single real number representing the limit of $\partial\cvx/\partial\cxd$ \emph{outside} the cusp.
Section \ref{section:vy} derives a model for the cusp shape as $\cy\eqs\cxd^\yex$, predicting cusp exponents $\yex$ which are also functions of $\mv,\wec$ alone 
(see \eqref{eq:correct-yex}). 

In section \ref{section:numerics} we test our model against numerical data. 
Section \ref{section:nummeth} describes the numerics used for calculating cusps and the new obstacles compared to previous sheet calculations. 
Section \ref{section:results} compares piecewise linear reconstruction to quadratic and higher degree; 
while the latter produce matching results that agree with our preferred model, 
we find that linear reconstruction produce a cusp exponent matching oversimplified conservation-violating models. 
(This is in contrast to the case of vortex spirals for which even point vortex methods were successful \cite{pullin-jfluid-1978,pullin-nuq}.)
Section \ref{section:xyexponent} and \ref{section:gxexponent} confirm that our $\gex,\yex$ formulas agree with numerical observations, 
with correct dependence on $\mv$ and $\wec$. 
Section \ref{section:clockwise} shows that clockwise circulation (on the upper sheet for a cusp opening to the right) is sometimes possible if $\wec$ is sufficiently large,
although generally counterclockwise circulation is observed, which matters because that is the circulation produced by Mach reflections 
(see \cite{neumann-1943,henderson-menikoff,serre-hbfluidmech,elling-tripleshock}). 
In cases where cusps do not appear we generally observe spiral-ends jets instead, as in section \ref{section:jetting} 
where we find that even for parameters far from the jet-cusp transition the jets can be rather small, mimicking cusps, 
suggesting that jetting is frequently obscured by viscous or kinetic effects in numerics or experiments, or simply small enough to be overlooked.

We conclude that our model is correct, that vortex cusps exist and can be computed accurately using vortex segments of at least quadratic degree.

\section{Modelling}
\label{section:modelling}

\subsection{Birkhoff-Rott equation}

\begin{figure}
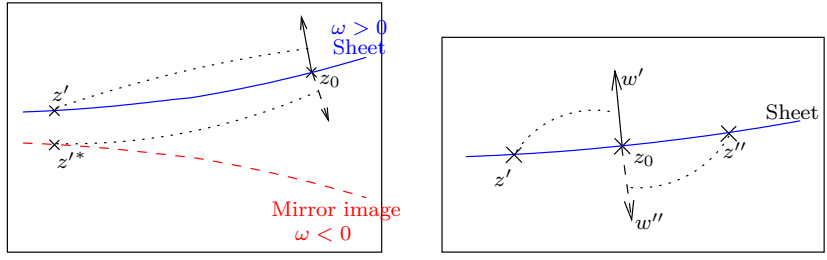

  \hfil\input{cancelposneg.pstex_t}\hfil\input{cancelnearpv.pstex_t}\hfil%
  \caption{Left: a point $\cz'$ on the sheet and its mirror image ${\cz'}^*$ induce almost opposite velocities in a distant point $\cz$.
    Right: two points near $\cz$ and at equal distances induce almost opposite velocities.}%
  \label{fig:cancel}%
\end{figure}

We are interested in a vortex sheet in the upper halfplane meeting the wall in a ``cusp''. 
We may assume the sheet approaches the cusp point from the right 
since approach from the left reduces to this case by mirror reflection. 
Equivalently we may think of a pair of vortex sheets in the whole plane (fig.\ \ref{fig:cancel}), mirror-symmetric across the horizontal axis, one above one below, 
with circulation of opposite sign. 
$\tgam$ is the circulation on the upper sheet, $\tz(t,\tgam)$ with $\tz=\tx+i\ty$ the location of the corresponding sheet point at time $t$. 
The infinitesimal segment $\tgam'$ to $\tgam'+d\tgam'$ of the upper sheet, regarded as an infinitesimal point vortex, 
induces in another point $\tz_0$ the complex velocity 
\begin{alignat}{5} \tvx-i\tvy = \frac1{2\pi i} \frac{d\tgam'}{\tz_0-\tz(t,\tgam')} . \notag\end{alignat} 
  The entire upper sheet induces a complex velocity
\begin{alignat}{5}  w(\tz_0) = \pvint \frac{d\tgam'}{2\pi i(\tz_0-\tz(t,\tgam'))} , \notag\end{alignat} 
where ``\text{p.v.}'' indicates principal value, the arithmetic average of the velocity limits on each side, i.e.\ of $w(\tz_\pm)$ as $\tz_\pm$ approach $\tz_0$ from each side of the sheet.
The lower sheet induces
\begin{alignat}{5} \frac1{2\pi i} \int \frac{-d\tgam'}{\tz_0-\tz^*(t,\tgam')} , \notag\end{alignat} 
where $-$ before $d\tgam'$ is due to  lower-sheet circulation of opposite sign,
whereas $\tz^*$ is the complex conjugate, the mirror image $\tx-i\ty$ of $\tz=\tx+i\ty$. 
The upper sheet evolves according to the \defm{Birkhoff-Rott equation}
\begin{alignat}{5} \tz_t(t,\tgam) = \twv^* \csep \twv = \pvint \frac{1}{\tz(t,\tgam)-\tz(t,\tgam')} - \frac{1}{\tz(t,\tgam)-\tz^*(t,\tgam')}\frac{d\tgam'}{2\pi i} \notag\end{alignat} 
where subscripts indicate partial derivatives. 
The integration domain is chosen based on the case at hand.
The Birkhoff-Rott equation is equivalent to the incompressible Euler equations under mild assumptions (see e.g\ \cite{lopes-lopes-schochet}). 

We are interested in \defm{self-similar} vortex sheets: since circulation $\tgam$ has dimensions of length squared over time, 
it is scaled by a factor $t^{2\mv-1}$. We pass to dimensionless $\cz,\cgam$ with the ansatz
\begin{alignat}{5} \tz(t,\tgam) = t^\mv \cz(1,t^{1-2\mv}\cgam) , \notag\end{alignat} 
resulting in the \defm{self-similar Birkhoff-Rott equation} 
\begin{alignat}{5} 0 = \cwv^* - \mv \cz + (2\mv-1)\cgam \cz_\cgam  \csep \cwv 
= \pvint 
\frac{1}{\cz(\cgam)-\cz(\cgam')} - \frac{1}{\cz(\cgam)-\cz^*(\cgam')} 
\frac{d\cgam'}{2\pi i}. 
\label{eq:ssbr}
\end{alignat} 
$\cwv^*-\mv\cz$ is the complex form of \defm{pseudo-velocity} $\cvq=\cvv-\mv\cxx$. 
$\cgam\mapsto\cz(\cgam)$ parametrizes the upper sheet, so $\cz_\cgam=\cx_\cgam+i\cy_\cgam$ is a tangent. 
Hence by \eqref{eq:ssbr} the pseudo-velocity $\cwv^*-\mv\cz=\cqx+i\cqy$ is everywhere tangent. 
This represents the pseudo-velocity \emph{on} the sheet; 
the limits \emph{on each side} of the sheet differ by adding or subtracting $\half\cgam_s\vec s$
where $\vec s$ is a sheet unit tangent, subscript $s$ indicates derivative with respect to arc length; hence the one-sided limits of pseudo-velocity are also tangential. 
$\cvq\dotp\vn=0$ corresponds to $\cvv\dotp\vn=\sigma$ for general unsteady vortex sheets, $\sigma$ being normal speed of the sheet.

When considering flows in the entire 2d plane we assume the velocity $\cwv$ is bounded or at least $o(|\cz|)$ near infinity (as $|\cz|\conv\infty$), so that velocity is uniquely determined by vorticity.
If so, then the term $-\mv\cz$ in \eqref{eq:ssbr} dominates $\cwv$ near infinity, leading to an approximation
\begin{alignat}{5} 0 = (2\mv-1)\cgam\cz_\cgam - \mv\cz \end{alignat} 
which has a simple solution
\begin{alignat}{5} \cgam = \const \cdot |\cz|^{2-1/\mv} \csep \frac{\cz}{|\cz|}=\const. \label{eq:outersol}\end{alignat} 
Hence the vortex sheets become asymptotic to straight lines near infinity. 
Those asymptotes correspond precisely to the initial data (fig. \ref{fig:Nsymm} left) in original $t,x$ coordinates: $t\dnconv 0$ there corresponds in the similarity plane to zooming out, $t\upconv\infty$ to zooming into the origin.

\subsection{Horizontal velocity approximation}
\label{section:vx}%

\subsubsection{Cusp contributions}

We consider cusp solutions of \eqref{eq:ssbr}. A key idea (see fig. \ref{fig:cancel} left) is that 
the velocity induced in a point $\cz_0$ by a not too close point $\cz'$ is almost completely cancelled 
by the mirror image ${\cz'}^*$. 
More precisely, an infinitesimal segment $d\cgam'$ at $\cz'$ on the upper sheet induces
\begin{alignat}{5} \frac{d\cgam'}{2\pi i (\cz_0-\cx'-i\cy')}  \label{eq:upperpoint} \end{alignat}
whereas its mirror image induces
\begin{alignat}{5} \frac{-d\cgam'}{2\pi i (\cz_0-\cx'+i\cy')}; \label{eq:lowerpoint} \end{alignat}
the sum is 
\begin{alignat}{5} 
d\cwv 
&= 
\frac{d\cgam} \pi \frac{\cy' }{ (\cz_0-\cx')^2+{\cy'}^2 } .
\label{eq:combinedw}
\end{alignat} 
$\cy'$ is very small near the cusp, so the sum is much smaller than the summands \emph{unless} $\cz_0$ is very close to $\cz'$. 

We emphasize the importance of this cancellation: the velocity integral, an operator that is generally non-local and thus complicated, 
is in the cusp case dominated by its \emph{local} part.
Of course ``local'' nurtures hopes of ``differential'', and indeed this convenient reduction will soon become apparent.

In addition, for $\cz_0$ on the upper sheet there is cancellation between the ``near'' parts to the left and to the right as well (see fig.\ \ref{fig:cancel} right): 
for positive $\cgam_\cx$ the left part induces almost upward velocities and the right one almost downward ones. 
So the dominant contribution to velocity is from the \emph{nearby mirror image}. 

\begin{figure}
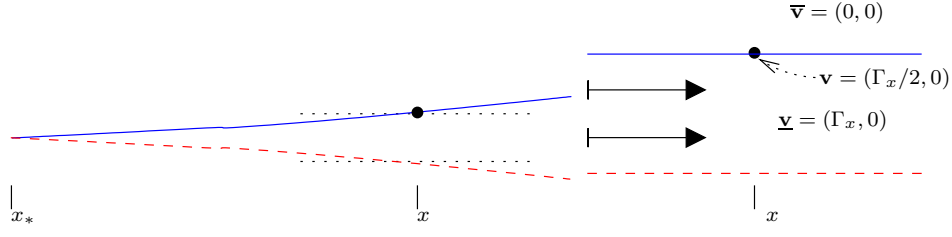

  \hfil\input{vapprox.pstex_t}\hfil\input{vapprox2.pstex_t}\hfil%
  \caption{Left: sheet tangents become almost horizontal near the cusp; right: velocity field for exactly horizontal sheets with constant velocity jumps $\pm\cgam_\cx$.}
  \label{fig:vapprox}
\end{figure}

To calculate its effect the following argument is convenient:
in near-cusp points $\cx+i\cy$ the sheet tangent is almost horizontal (fig.\ \ref{fig:vapprox} left);
hence we approximate the sheet and its mirror image by two straight lines 
with constant velocity jump $\cgam_\cx(\cx)$ (fig. \ref{fig:vapprox} right).
Such an approximation has velocity $\cva=(0,0)$ above the upper sheet (and below the lower one), $\cvi=(\cgam_x,0)$ between the two sheets; on the upper sheet the principal value is the arithmetic average $\cvs=(\half\cgam_x,0)$. For the real (horizontal) part of \eqref{eq:ssbr}:
\begin{alignat}{5} 0 &= \cvx - \mv\cx + (2\mv-1)\cgam\cx_\cgam    \notag\\
  &\approx  \half\cgam_\cx - \mv\cx + (2\mv-1)\cgam \cx_\cgam .  \label{eq:cvxwithout} \end{alignat} 
In one case our approximations are appropriate not only near the cusp point but everywhere: 
let $2\phiinf$ be the angle the straight-line asymptotes of vortex sheet and mirror image enclose near infinity (fig. \ref{fig:Nsymm} left). 
If the \defm{infinity angle} $\phiinf$ is \emph{small}, then we may expect that the sheet tangents are nearly horizontal everywhere.

\subsubsection{Non-cusp contributions; symmetry}

Of course we have neglected that there are velocity contributions from the outer (non-cusp) part of the sheets, 
as well as those of other sheets, vortex patches or any other form of vorticity present in the 2d similarity plane. 
Since they are uniformly distant from the cusp we may model the velocity induced by them as a complex-analytic function $\wve=\wve(\cz)$. 
We use its Taylor expansion
\[ \wve(\cz) = \wez + \wec (\cz-\cxu) + o(|\cz-\cxu|)  \] 
where $\cxu$ is the cusp location which is real (horizontal axis); 
the coefficients $\wez,\wec$ are likewise real since velocity is horizontal at the wall by the slip condition. 
In section \ref{section:e-depth} we argue that expansions beyond $\wec$ cannot always be justified.

The constant $\wez$ does not have a direct influence on the cusp shape; it merely corresponds to a constant shift in the similarity plane.
The \defm{strain coefficient} $\wec$ corresponds to a saddle flow $\cvv=(\wec(\cx-\cxu),-\wec\cy)$; for $\wec>0$ this flow is 
expanding along the symmetry axis of the cusp (horizontal axis), compressing in the perpendicular direction. 

The horizontal part of $\wve$ is 
\[ \Re\wve = \wez + \wec \cxd + ... \] 
where $\cxd=\cx-\cxu$ is horizontal distance to the cusp; 
we neglect (at this point) the vertical part
\begin{alignat}{5} \Im\wve(\cz) = \wec\cy + ... \label{eq:Ime} \end{alignat}
since $\cy$ decays rapidly in a cusp. 
Now \eqref{eq:cvxwithout} is replaced by the more accurate model
\begin{alignat}{5} 0 = \half\cgam_\cx + \subeq{\wez + \wec\cxd}{\approx\Re\wve}  - \mv\cx + (2\mv-1)\cgam \cx_\cgam \label{eq:cvxwith} \end{alignat} 

In a few special cases we may make additional assumptions. 
E.g.\ for small infinity angle $\phiinf$, and without non-cusp flow features present, we may assume that 
$\wve$ becomes small (possibly in contrast to $\half\cgam_\cx$). 

Another special case is $\Nv$-fold symmetry (see fig.\ \ref{fig:Nsymm}): 
instead of a single pair of equal-strength opposite-circulation sheets consider $\Nv$ pairs, 
with equal angles $2\pi/\Nv$ between the symmetry axes of adjacent pairs, 
and cusps in the origin (as suggested by numerical experiments). 
Then by symmetry the Taylor expansion of $\wve$ vanishes up to and including $\cz^{\Nv-1}$. 
For $\Nv=2$ that means $\wez=0$; for $\Nv\geq 3$ we additionally have $\wec=0$. 
Symmetry is a convenient device for suppressing $\wez,\wec$ so that other influences shaping the cusp, 
such as dependence on $\mv$ or $\phiinf$, can be studied in isolation.

Symmetry also helps avoiding problems at infinity: 
the integral in the self-similar Birkhoff-Rott equation \eqref{eq:ssbr}
behaves like $\int|\cz|^{-1/\mv}d|\cz|$ near infinity which generally diverges for any $\mv\geq 1$. 
In special cases it does converge; for instance with $\Nv\geq 2$ symmetry the antipodal vortex sheets 
cancel out sufficiently to improve the integrand to $\eqs|\cz|^{-1-1/\mv}$,
convergent for all $\mv<\infty$. Even so infinity becomes dominant like $O(\mv)$ as $\mv\upconv\infty$,
but for $\Nv\geq 3$ a further cancellation avoids even that effect.

\subsubsection{Conservation arguments}
\label{section:consmass}%

\begin{figure}
  \centerline{\input{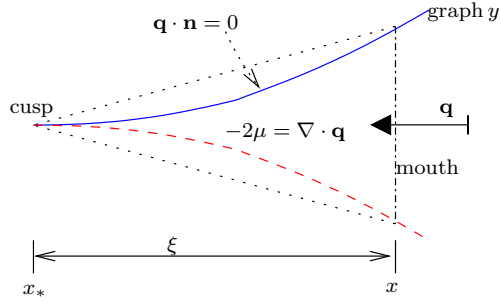}}
  \caption{Mass conservation implies negative $\cqx$ averages over the right side of cusp triangles}
  \label{fig:cusptriangle}
\end{figure}
\eqref{eq:cvxwith} is not only a differential equation but an ordinary one; 
we have modelled $\vxs$ as a function of $\cxd$ alone. 
The ODE is not difficult to solve, but we minimize effort on irrelevant solutions by first considering 
conservation of mass. $\ndiv\vv=0$ has a self-similar form
\begin{alignat}{5} \ndiv\cvq = \ndiv(\cvv-\mv\cxx) = -2\mv  \notag\end{alignat} 
where $2$ comes from the number of dimensions. Integrate this over a small ``cusp triangle'' (formed in fig.\ \ref{fig:cusptriangle} by solid and dashed vortex sheet and dashed-dotted ``mouth'')
from the cusp $\cxu$ to some $\cx>\cxu$:
since the normal pseudovelocity $\cvq\dotp\vn$ is zero at each side of a vortex sheet, only the ``mouth'' boundary term remains:
\begin{alignat}{5} \int_{-y(\cxd)}^{y(\cxd)} \cqx ds = -2\mv \cdot \text{cusp triangle area} \notag\end{alignat} 
On the left we use $\cqx\approx\qxi(\cxd)$. The right-hand side is always negative, but 
for a ``cusp'' the area is reasonably estimated as less than the area of the straightened-side triangle 
(dotted in fig.\ \ref{fig:cusptriangle}) which is $\cxd y(\cxd)$. Hence
\begin{alignat}{5} 0 > \qxi \geq - \mv \cxd , \label{eq:qxibound}\end{alignat} 
and as $\cxd$ is increased $\qxi$ is strictly decreasing.
Velocity $\cvx$ must be a constant plus $O(\cxd)$, and same for $\cgam_x$ which differs from $\cvx$ by a smooth function $\Re\wve$.

\subsubsection{Constraints}
\label{section:constraints}

Multiply both sides of \eqref{eq:cvxwith} by $\cgam_\cx$ to obtain 
\begin{alignat}{5} 0 
  &= \half\cgam_\cx^2 + (\wez+\wec\cxd - \mv\cx)\cgam_\cx + (2\mv-1)\cgam;  \notag
\end{alignat}
completing the square for $\cgam_\cx$ yields 
\begin{alignat}{5}
  0 &= \half(\cgam_\cx + \wez+\wec\cxd - \mv\cx )^2 -\half (\wez+\wec\cxd-\mv\cx)^2 + (2\mv-1)\cgam.  \notag
\end{alignat}
The first parenthesis is recognized to be $\qxi$ since (recall fig.\ \ref{fig:vapprox} right) the self-induced velocity is $\cgam_\cx$ between the sheets, so 
\begin{alignat}{5}
 \qxi &= \vxi - \mv\cx = \cgam_\cxd + \wez + \wec\cxd - \mv\subeq{\cx}{=\cxu+\cxd}. \label{eq:gxqpre}
\end{alignat}
So we obtain
\begin{alignat}{5}
0 = \half(\qxi)^2 - \half(\wez+\wec\cxd-\mv\cx)^2 + (2\mv-1)\cgam . \label{eq:qxig}
\end{alignat} 
Mass conservation was shown to require $\qxi\upconv 0$ as we approach the cusp ($\cxd\dnconv 0$), which implies (for $*$ indicating values in the cusp limit)
\begin{alignat}{5} (2\mv-1)\cgam_* = \half(\wez-\mv\cxu)^2 \geq 0  \label{eq:cgse}
\end{alignat} 
Hence $\cgam_*=0$ or $\cgam_*>0$; the latter appears to happen only in borderline cases, so we defer it to section \ref{section:xcusps}.

\subsubsection{$\cx$ ODE solutions in case $\cgam_*=0$}
\label{section:cx-solve}

By \eqref{eq:cgse} $\cgam_*=0$ requires 
\begin{alignat}{5} \wez = \mv\cxu , \label{eq:cuspwve}
\end{alignat} 
so \eqref{eq:gxqpre} and \eqref{eq:qxig} simplify to 
\begin{alignat}{5}
  \qxi &= \cgam_\cxd + (\wec-\mv)\cxd \quad[\quad\eqv\quad \cgam_\cxd = \qxi+(\mv-\wec)\cxd \quad] \label{eq:gxq} 
\end{alignat}
and
\begin{alignat}{5}
0 &= \half(\qxi)^2 - \half((\wec-\mv)\cxd)^2 + (2\mv-1)\cgam .\notag
\end{alignat}
After $\pd\cxd$ of the latter, producing
\begin{alignat}{5}
  0 = \qxi\qxi_\cxd - (\mv-\wec)^2\cxd + (2\mv-1)\cgam_\cxd, \notag
\end{alignat}
the former substitutes the final $\cgam_\cxd$ to yield
\begin{alignat}{5}
  0 
  &= \qxi\qxi_\cxd +(\mv-1+\wec)(\mv-\wec)\cxd + (2\mv-1)\qxi . \label{eq:qxiz}
\end{alignat}
This ODE turns autonomous if we eliminate the $\cxd$ dependence by the substitution
\begin{alignat}{5}
  \qxi(\cxd)=\cxd Q(\cxd)  \csep  \cgam_\cxd(\cxd) = \cxd G(\cxd)  \label{eq:qQ}
\end{alignat}
(which is also natural in light of $\qxi=O(\cxd)$) to obtain
\begin{alignat}{5}
  0 
  &= \cxd^2 QQ_\cxd+\cxd Q^2 +(\mv-1+\wec)(\mv-\wec)\cxd + (2\mv-1)\cxd Q .
\end{alignat}
Divide by $\cxd>0$ and factor using $(\mv-1+\wec)+(\mv-\wec)=2\mv-1$:
\begin{alignat}{5}
  \frac{\partial Q}{\partial(-\log\cxd)} 
  &= 
  \frac{ -(Q+\mv-\wec)(Q+\mv+\wec-1) }{-Q}.
  \label{eq:Qeq}
\end{alignat}
Our conservation-of-mass arguments showed that $Q=\qxi/\cxd$ is negative but bounded in the cusp limit $-\log\cxd\conv+\infty$ (see \eqref{eq:qxibound}).
This is possible only if $Q$ converges to a nonpositive root of the numerator. The two roots are
\begin{alignat}{5}
  Q &= 1-\wec-\mv \quad\topref{eq:gxq}{\rightsquigarrow}\quad G =  1-2\wec, \label{eq:root1} \\
  Q &= \wec-\mv \quad\topref{eq:gxq}{\rightsquigarrow}\quad G = 0 . \label{eq:root2}
\end{alignat}
The two roots are equal if and only if $\wec=\half$, a borderline case we ignore to keep the discussion concise and avoid $\log\cxd$ terms.

First consider solutions that converge to one of the roots from one side. Then that root must be asymptotically stable on that side. 
The right-hand side of \eqref{eq:Qeq} has positive denominator for $Q<0$ whereas the leading $-Q^2$ part of the quadratic numerator makes it positive between the roots, 
negative outside. Hence if the roots are distinct, the smaller root is either unstable on both sides or not negative and hence irrelevant, 
so we may ignore it.
On the other hand if the larger root is negative, then it is asymptotically stable on both sides. 
The corresponding solutions will indeed turn out to lead to meaningful vortex cusps. 

If $\wec<\half$, then $Q=1-\wec-\mv$ root is the larger root.
It is negative if and only if $\wec>1-\mv$. We only consider $\mv>\half$ (see introduction), so $1-\mv<\half$, 
meaning the range $\boi{1-\mv}{\half}$ for $\wec$ is never empty. However, the constraint $1-\mv<\wec$ is significant especially for smaller $\mv$; 
for $\mv<1$ it does not permit cusps with $\wec$ arbitrarily small.
The root corresponds to $G = 1-2\wec+...$, so we obtain solutions
\begin{alignat}{5}
  \cgam_\cxd &= (1-2\wec)\cxd + o(\cxd) \csep \cgam = (\half-\wec)\cxd^2 + o(\cxd^2). \label{eq:gx-prestandard}
\end{alignat}

If on the other hand $\wec>\half$, then $Q=\wec-\mv$ is the larger root. It is negative if and only if $\wec<\mv$; 
again the range $\boi{\half}{\mv}$ is nonempty but especially for smaller $\mv$ a significant constraint.
The root corresponds to $G=0+...$, so the leading term of $\cgam$ has to be determined by linearizing \eqref{eq:Qeq} there: passing to 
\begin{alignat}{5}
  \frac{\partial G}{\partial(-\log\cxd)} \topref{eq:Qeq}{=} 
  \frac{ -G(G+2\wec-1) }{-G+\mv-\wec}
  \notag 
\end{alignat}
the linearization at $G=0$ is 
\begin{alignat}{5}
  \frac{\partial G}{\partial(-\log\cxd)} 
  =
  -\frac{2\wec-1}{\mv-\wec} G
  \notag 
\end{alignat}
with solutions $G\eqs\cxd^{(2\wec-1)/(\mv-\wec)}$.

In summary:
\begin{alignat}{5}
  &\cgam_\cxd = C\cxd^{\gex-1} + ... \csep \cgam = \frac{C}{\gex} \cxd^\gex + ...,  \label{eq:gx-nonstandard}\\
  &\fbox{$\displaystyle \gex = \begin{cases}
    2 , & 1-\mv < \wec < \half, \\
    2 + \frac{2\wec-1}{\mv-\wec} ,  & \half < \wec < \mv ,
  \end{cases} $} \label{eq:gex}\\
  &C = \begin{cases} 1-2\wec , & \wec<\half \\ \text{free}, & \wec>\half .\end{cases} \label{eq:gxconst}
\end{alignat}

Finally consider solutions that \emph{equal} one of the roots, so that asymptotic stability is not necessary. 
$Q=\wec-\mv$ corresponds to $\cgam_\cx=0$, the trivial case of no sheets at all which we ignore. 
$Q=1-\wec-\mv$ for $\wec<\half$ is a special case of the discussion above;
for $\wec>\half$ it does not produce cusps, as will become apparent at the end of section \ref{section:correct}.

\subsection{Vertical velocity modelling}
\label{section:vy}%

\subsubsection{Simple model}

Given the relationship between $\cxd$ and $\cgam$, it remains to derive a model for the upper sheet height $\cy=\cy(\cxd)$, using 
the imaginary (vertical) part of the self-similar Birkhoff-Rott equation \eqref{eq:ssbr}:
\begin{alignat}{5} (1-2\mv)\cgam\cy_\cgam + \mv\cy = \cvy . \label{eq:ssbry}\end{alignat} 
Returning to our idea for approximating the lower and upper sheet as locally straight and horizontal (fig.\ \ref{fig:vapprox}), 
the simplest model is to adopt the resulting cusp-induced velocity $\cvy=0$. Together with the non-cusp part $-\Im\wve$ from \eqref{eq:Ime}
we obtain
\begin{alignat}{5} (1-2\mv) \frac{\cgam}{\cgam_\cxd} \cy_\cxd + \mv \cy = 0 - \wec\cy. \notag  \end{alignat} 
\eqref{eq:gx-nonstandard} yields $\cgam/\cgam_\cxd=\cxd/\gex$,
so near the cusp the approximation
\begin{alignat}{5} \frac{1-2\mv}{\gex} \cxd \cy_\cxd = (-\wec- \mv)\cy  \notag \end{alignat} 
follows, with solution
\begin{alignat}{5} \cy &= \const\cdot \cxd^\yex \notag \end{alignat}
for \defm{cusp exponent}
\begin{alignat}{5} \yex = \frac{\gex(\mv+\wec)}{2\mv-1}.  \notag \end{alignat}
Substitute \eqref{eq:gex} to obtain 
\begin{alignat}{5} \yex = \begin{cases} 
  \frac{2(\mv+\wec)}{2\mv-1} , &\wec < \half, \\
  \frac{\mv+\wec}{\mv-\wec} , &\wec > \half.
\end{cases} 
\label{eq:simple-yex}\end{alignat}

\subsubsection{Improved model}

\begin{figure}
  \centerline{\input{improved.pstex_t}}
  \caption{Improved $\cvy$ approximation: $[\cvv\dotp\vn]=0$ and $[\cvx]$ determine $[\cvy]$.}
  \label{fig:improved}
\end{figure}

However, we will see later that \eqref{eq:simple-yex} is not the correct exponent if $\wec<\half$. 
While the horizontal-straight-sheet approximation is appropriate for the $\cx$ ODE, 
$\cy$ decays faster so that relative errors are larger. 

A better approximation is obtained by the following simple argument 
(fig.\ \ref{fig:improved}): the jump of velocity from below to above the upper sheet is \emph{tangential},
hence must be 
\[ -\frac{\cgam_s}{\sqrt{1+\cy_\cx^2}}\begin{bmatrix}1\\\cy_\cx\end{bmatrix} \] 
($-$ due to counterclockwise circulation for positive $\cgam_s$). 
We allow nonzero but still small slopes $\cy_\cx$, so to leading order the square root is $1$ and also $-\cgam_s\approx-\cgam_\cx$. 
So we still obtain the same approximation $-\cgam_\cx$ for the $\cvx$ jump, but now $\cvy$ jumps by $-\cgam_\cx\cy_\cx$. 
To respect the slip condition we take $\vyi=0$ on the lower side of the upper sheet, yielding $\vya=-\cgam_\cxd\cy_\cxd$ on the upper side and an arithmetic average
\begin{alignat}{5} \vys = -\half\cgam_\cxd\cy_\cxd \notag\end{alignat} 
on the sheet. Now \eqref{eq:ssbry} yields the improved model 
\begin{alignat}{5} \frac{1-2\mv}{\gex} \cxd \cy_\cxd + \mv \cy = -\half\cgam_\cxd\cy_\cxd - \wec\cy. \notag\end{alignat} 
Compare the new $\cy_\cxd$ term on the right-hand side to the left-hand side one which has coefficient $\eqs\cxd$. 
For $\wec<\half$ the new coefficient $\cgam_\cxd$ is $\eqs\cxd$, hence its term non-negligible, and \eqref{eq:gxconst} yields
\begin{alignat}{5} \frac{1-2\mv}{2} \cxd \cy_\cxd + \mv \cy = -\half(1-2\wec)\cxd\cy_\cxd - \wec\cy \notag\end{alignat} 
with solution
\begin{alignat}{5} \cy &= \const\cdot \cxd^\yex \csep \yex 
  = \frac{\mv+\wec}{\mv-1+\wec}  \label{eq:improved-yex}\end{alignat} 
That this ``improved'' exponent is not the ``simple'' exponent in \eqref{eq:simple-yex} already confirms that our previous approximation 
omitted a significant term.

\subsubsection{Correct model}
\label{section:correct}

Both numerical experiments and heuristic arguments will show later that the improved exponent $\yex$ is \emph{still} incorrect.
The simplest way to see the inadequacy is to invoke again the conservation of mass arguments from section \ref{section:consmass}. 
Again integration of $-2\mv=\nabla_\cxx\dotp\cvq$ over the cusp triangle (fig.\ \ref{fig:cusptriangle}) yields
\begin{alignat}{5} -2\mv \int_0^\cxd 2\cy(\cxd)d\cxd = 2\cy(\cxd) \qxi(\cxd) + ... ; \end{alignat} 
omitting all but the leading terms and using $\cy(\cxd)=C\cxd^\yex+...$ and 
the $\qxi$ asymptotics from \eqref{eq:gxq} and \eqref{eq:gx-nonstandard} we see
\begin{alignat}{5} \frac{-2\mv}{\yex+1} 2C\cxd^{\yex+1} = 2C\cxd^{\yex+1} \cdot \begin{cases} 1-\wec-\mv , & \wec<\half, \\ \wec-\mv, & \wec>\half. \end{cases} \end{alignat} 
We solve for 
\begin{alignat}{5} 
  \fbox{$\displaystyle\yex 
    = \begin{cases} \frac{\mv+1-\wec}{\mv+\wec-1} , & 1-\mv<\wec<\half, \\ \frac{\mv+\wec}{\mv-\wec} , & \half < \wec < \mv.  \end{cases} $}
  \label{eq:correct-yex} 
\end{alignat}
Hence, given $\cgam(\cxd)$, conservation of mass already determines $\cy(\cxd)$!

We emphasize that $\wec$ is not a fudge factor that can be used to realize any observed cusp exponent $\yex$. 
For example in case of $\Nv\geq 3$ symmetry $\wec$ is zero so that the prediction \eqref{eq:correct-yex} depends only on $\mv$. 
Even without symmetry, $\wec$ can be measured independently, for example in numerics by calculating the velocity field derivative 
in an evaluation point near the cusp but outside and not too close so that the cusp part of the sheet pair almost cancels. 
Besides, the $\cgam,\cx$ relationship can also be measured; the formula \eqref{eq:gex} for $\gex$ is likewise dependent
on $\wec$ and $\mv$, providing a second constraint for the two. 
$\wec$ depends on the vorticity in \emph{every} part of the similarity plane; it cannot be modelled from local considerations.

Our restrictions
\begin{alignat*}{5}
  1-\mv<\wec<\mv
\end{alignat*}
have physical interpretations. The lower bound is due to $\qxi<0$ inside the cusp. 
Upper bound: consider a flow with no vortex sheets at all, but a velocity $\wve=\wec\cz$. 
Then the pseudo-velocity $\cvq=\cvv-\mv\cxx=((\wec-\mv)\cx,(-\wec-\mv)\cy)$ has a zero $\cxx=0$
that is a sink if $\wec<\mv$, 
but a \emph{saddle} if $\wec>\mv$. Naturally a background pseudo-velocity that is expanding along the horizontal axis 
does not favor sheets \emph{approaching} the origin along that axis, let alone forming cusps, 
unless the sheets are sufficiently strong to overcome the background with their self-induced velocity. 

At this point we may collect the loose end left in section \ref{section:cx-solve}
where we ignored the constant solution $Q=1-\wec-\mv$, i.e.\ $\cgam_\cx=(1-2\wec)\cx$, in case $\wec>\half$. 
Repeating our conservation of mass argument we obtain
\begin{alignat*}{5}  \yex = \frac{\mv+1-\wec}{\mv+\wec-1} \overset{\wec>\half}{\leq} 1 . \end{alignat*}
But a proper $\cy\eqs\cx^\yex$ cusp, in particular our small-slopes assumption $|\cy_\cx|\ll 1$, requires $\yex>1$.
Although this $Q$ represents an appropriate solution of the $\cx$ ODE, it leads to an unusable solution of the $\cy$ ODE. 
Hence we were justified in rejecting it. 

A full ``correct'' model ODE yielding $\yex$ as in \eqref{eq:correct-yex} can be derived by
considering again conservation in the cusp triangle (fig.\ \ref{fig:cusptriangle}):
\begin{alignat*}{5}
  -2 \mv \cdot \text{area} &= 2 \cy \qxi
\end{alignat*}
Take $\pd\cxd$, corresponding to moving the ``mouth'':
\begin{alignat*}{5}
  -2 \mv \cdot 2 \cy &= 2 \cy_\cxd \qxi + 2 \cy \qxi_\cxd
\end{alignat*}
Solve for 
\begin{alignat}{5}
  \frac{\cy_\cxd}{\cy} &= \frac{2\mv+\qxi_\cxd}{-\qxi} \label{eq:correct-yode-qxi}
\end{alignat}
Substitute \eqref{eq:gxq}:
\begin{alignat}{5} \cxd \frac{\cy_\cxd}{\cy} 
  &= 
  \frac{\cgam_{\cxd\cxd}+\wec+\mv}{\mv-\wec-\cgam_\cxd/\cxd}.
  \label{eq:correct-yode}\end{alignat} 
For $\wec<\half$ \eqref{eq:gx-nonstandard} shows
\begin{alignat}{5} \cxd \frac{\cy_\cxd}{\cy} 
  &= 
  \frac{(1-2\wec)+\wec+\mv+...}{\mv-\wec-(1-2\wec)+...}
\end{alignat} 
which generates the desired exponent.
The ODE can also be used to obtain terms beyond the leading ones, but as the asymptotic expansion deepens
physical effects we neglected during derivation of the model may become significant.

\section{Numerical validation}
\label{section:numerics}

\subsection{Methods}
\label{section:nummeth}

\begin{figure}
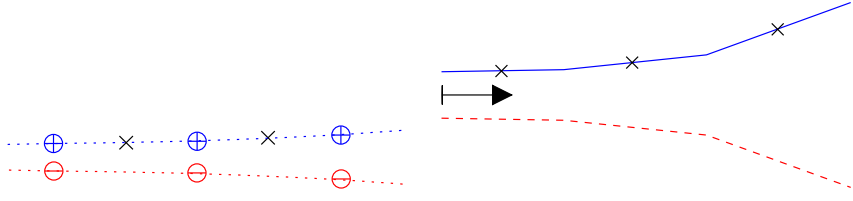

  \hfil\input{pointvortex.pstex_t}\hfil\input{linesegment.pstex_t}\hfil%
  \caption{Left: point vortex discretization of sheet pair; velocities are singular near each $\pm$ pair, near-zero elsewhere. 
    Right: piecewise linear; $\cvx$ is sufficiently accurate in the evaluation points (marked $\times$), but $\cvy$ is not.}
  \label{fig:pointvortex}
\end{figure}

To test our models we compare their predictions to numerical data obtained from discretizing the Birkhoff-Rott equation itself. 
Besides general methods for unsteady vortex sheet rollup or other free-surface evolutions (e.g.\ \cite{krasny-icm} or \cite{meiron-baker-orszag}),
numerics adapted to self-similar flow using point vortices were successfully used to compute algebraic vortex spirals in \cite{pullin-jfluid-1978,pullin-nuq}. 
The vortex sheet was approximated by replacing the continuous sheet 
with discrete point vortices, choosing evaluation points for the Birkhoff-Rott equation midway between adjacent vortices
to mitigate the strong singularity of the point vortex velocity field. 

In our experience this approach is quite effective for vortex spiral calculations, well beyond the cases discussed by \cite{pullin-jfluid-1978,pullin-nuq}. 
But in the case of vortex \emph{cusps} or similar cases it is clearly ineffective,
due to the cancellation between positive and negative vorticity 
(see fig.\ \ref{fig:pointvortex} left): if the upper sheet is resolved with a moderate number of point vortices they will near the cusp 
inevitably be at wide horizontal spacing relative to the vertical distance to the mirror vortex.
Then the velocity is highly singular in their vicinity, but nearly zero velocity away from each pair.
Clearly such a field has no resemblance to the real one. 
To obtain a reasonable approximation the horizontal spacing must be small compared to the sheet-image gap.
But this is rather expensive because in a cusp the sheet and mirror image approach very rapidly. 

Hence we must use low horizontal resolution and increase the degree instead. 
A natural concern is that physically relevant wavelengths may be suppressed, leading to spurious solutions. 
But oscillations are not expected in self-similar (as opposed to unsteady) vortex sheets; 
e.g.\ it has been shown that vortex sheet with some minimum regularity that exist for a positive time must a fortiori be analytic (\cite{lebeau}, \cite{wu-chordarc}). 
It is also easy, especially after a solution has already converged at coarse resolution, 
to test by selecting a few segments and refining them far more than the rest, to the point where horizontal resolution is much finer than the vertical gap; 
none of our experiments suggested that additional resolution was required. 

While point vortices have obvious problems, it is more surprising that piecewise linear vortex segments are \emph{still} inadequate. 
The strongest hint is in our ``correct'' $\cy$ ODE \eqref{eq:correct-yode}: 
it contains a term $\cgam_{\cxd\cxd}$ that cannot be resolved by a piecewise linear relationship between $\cxd$ and $\cgam$. 
Indeed our numerical results will show clearly that a continuous piecewise linear representation
produces cusps with the ``improved'' cusp exponent \eqref{eq:improved-yex}, not the ``correct'' one \eqref{eq:correct-yex}. 
In fact the linear approximation violates conservation of mass like the ``improved'' model did, 
as can be observed in numerical solutions by testing $-2\mv=\ndiv\cvq$ via numerically evaluated mass flux boundary integrals.
The defect is not cured by moderate refinement. 
It turns out that piecewise quadratic or higher approximation is needed to obtain accurate results. 

This is rather inconvenient: not only are higher-degree representations segment-by-segment more expensive to evaluate than linear ones, 
which are already more costly than point vortices, but in addition higher order numerics are prone to oscillations and stability problems.
Fortunately higher order is only needed for segments that are close to the evaluation point. 

In detail our approximation works as follows: we parametrize the upper sheet by a function $\cz$ of $\cgam$. 
The numerical scheme solves for variable $\cz_i$ approximating $\cz(\cgam_i)$ in finitely many $\cgam_i$. 
(For point vortices it is important that the $\cgam_i$ are evenly spaced, but as the degree of approximation increases the influence of spacing decreases.)
The sheet segment from $\cz_i$ to $\cz_{i+1}$ is approximated by a polynomial interpolating these two points and a few adjacent one, 
for example $\cz_{i+2}$ or $\cz_{i-1}$ for quadratic and both for cubic, etc. 

Although the approximated sheet is continuous, it may have corners at each $\cz_i$ even if the degree is high. 
It is of course possible to require tangents and possible higher derivatives to match, 
but this turned out to have unpleasant side effects in our experiments. 
Asking for continuity of as many derivatives as possible leads to natural cubic splines and similar interpolation schemes, which are non-local, slowing the calculation,
although locality can be obtained by compromising on top-order continuity. 
But more importantly the higher the spline smoothness the higher the tendency for oscillations at nonsmooth parts of the sheet.
Consider for example the fork point in fig.\ \ref{fig:cusp} second from top, or for a more extreme example the ODE solution for $\mv=1+\eps$
which is almost $\cy(\cxd)=0$ for $x$ inside the cusp, $\cy(\cxd)=\const\cdot\cxd$ outside, i.e.\ a kink). 
Trying to approximate such shapes by natural cubic splines would cause strong oscillations near the kink,
with upper-lower sheet intersection very hard to avoid, among other problems.

On the other hand if the sheet approximation has corners, then the velocity integral produces logarithmic singularities. 
In our experiments this did not appear to be important. 
As a sufficiently smooth part of the sheet is refined, the coefficient of each logarithmic singularity decreases. 
Moreover the evaluation points are chosen in the midpoints between the corners, far from the singularities 
and possibly with an extra order of cancellation,
similar to \cite{pullin-jfluid-1978,pullin-nuq} where point vortices had far more severe $1/r$ singularities in place of the mild $\log r$. 

For a piecewise polynomial function $\cz$ of $\cgam$ the Birkhoff-Rott integral $\int(Z-\cz)^{-1}d\gam$ can be done by partial fractions, yielding formulas with several complex logarithms. 
For higher degree this requires polynomial root finding. For the purpose of this paper, more concerned with correctness of the modelling rather than optimizing speed of numerical schemes,
we simply used the usual quadratic or Cardano formulas.

The Birkhoff-Rott equation \eqref{eq:ssbr} is discretized by evaluation in the arithmetic averages $\cgam_{i+\half}$ of $\cgam_i,\cgam_{i+1}$. 
On the left-hand sides $\cz$ and its derivatives are replaced by the segment $(i,i+1)$ polynomial approximation. 
This produces finitely many nonlinear algebraic equations with complex values depending on complex variables $\cz_i$. 
Newton iteration is applied to the system, solving the linearization by direct methods using the exact derivative matrix. 
Although the cubic cost growth limited the number of vortices to a few thousand, this was more than sufficient for our purposes so that it was not necessary
to explore more efficient alternatives. 

While for the exact sheet $\cgam$ ranges from $0$ (cusp) to $\infty$ (outer limit), the approximation uses inner and outer cutoffs. 
At the outer cutoff several fixed $\cz$ are calculated by the straight-sheet asymptotes \eqref{eq:outersol}. 
The outer cutoff must be taken large enough to avoid large errors, especially for $\Nv\leq 2$ and large $\mv$. 
At the inner cutoff, for the innermost segment $z_0,z_1$ the polynomial degree was limited to quadratic, using $z_2$ as third interpolation point. 

The innermost point and mirror image have a small nonzero gap. 
There the sheet and image having a tendency to split apart if the calculation becomes unstable (e.g.\ if overly large Newton iteration steps are taken
or if the initial approximation is poor). 
We experimented with extending the sheet by fitting $y\eqs\cxd^\yex$ cusp shapes and generating additional $\cz$ points on them, 
but stability seemed to worsen rather than improve.

Instead of calculating ``global'' cusps, especially for $\Nv=1$ (no symmetry) it is interesting to study ``local'' ones, with a small outer cutoff; 
this is reasonable since in many applications the vortex cusp is merely part of a larger incompressible flow, 
or a quasi-incompressible region of a compressible flow. Changing the outer cutoff adds vorticity at a uniform distance from the cusp, 
hence induces a holomorphic velocity contribution, so the effect of cutoff changes can be modelled by adjusting our $\wve$. 
In addition, to stabilize the solution against outer-sheet changes, it is possible to subtract from all calculated $\cwv$ the value computed in a fixed point, 
say the origin. A constant added to $\cwv$ can in \eqref{eq:ssbr} be absorbed by a constant shift added to $\cz$, since every term except $\mv\cz$ is either a difference of two $\cz$ or a derivative of $\cz$. 
Hence only the location of the cusp is changed, but not its shape or circulation distribution. 

We do not specify the \emph{strength} of the vortex sheets 
since it can be scaled arbitrarily by scaling of time $t$ and space $x$, corresponding to scaling the similarity coordinate $t^{-\mv}x$.
The sheet strength determines the scale at which the sheet pair transitions from near-straight lines to a cusp.
This scale could be defined as (to give only one example) the radius at which the cusp point angle between sheet points $\cz$ and their image $\cz^*$ has decreased from $2\phiinf$ to $\phiinf$ .

For calculating spiral-end jets we combine our new methods with the techniques of Pullin:
in spiral parts of sheets we approximate $\cgam$ and spiral center distance $r$ as functions of traversed angle $\beta$, 
choosing $\beta$ step width to be $2\pi$ divided by an integer, so that interpolation points lie on several common rays,
with evaluation points at half-angles between these rays. 
The transition between coordinates $\beta$ inside the spiral and $\cgam$ outside is troublesome, but less unstable than in pure point vortex calculations; adaptive refinement, which is easily implemented in 1d, is helpful in avoiding coordinate singularities. 

We emphasize that our observations leading to the correct ODE model apply to all flows with sheet and mirror image that are close and nearly parallel, even if they do not terminate in a joint cusp,
so for calculations of \emph{narrow} jets quadratic or higher degree is still essential, e.g.\ for the outside-spiral parts of fig.\ \ref{fig:cusp} bottom.

Like Pullin's point vortex calculations ours are very sensitive due to the inherent instability of vortex sheets. 
An added concern is the proximity of sheet and image; intersection must be avoided carefully, although for special cases it is possible to modify branch cuts in logarithms so that the approximate BR integral remains smooth
even as upper-sheet vortices pass below lower-sheet ones. (Of course such intersections are permissible only during iteration, not in a converged solution.)

It is important to start with a good initial approximation, such as those obtained from our ODE models, 
and to limit iteration step size until a good basin of attraction is reached. 
In particular at the inner cutoff the sheets are prone to separate, causing uncontrollable oscillations and self-intersection. 
For large infinity angles $\phiinf$ we found it helpful to start with numerical solutions for small ones, then gradually increasing the angle. 
This numerical homotopy was also useful for other parameters like $\mv$ or $\wec$.

\subsection{Results}
\label{section:results}

\begin{figure}
  \centerline{\includegraphics[width=.7\linewidth]{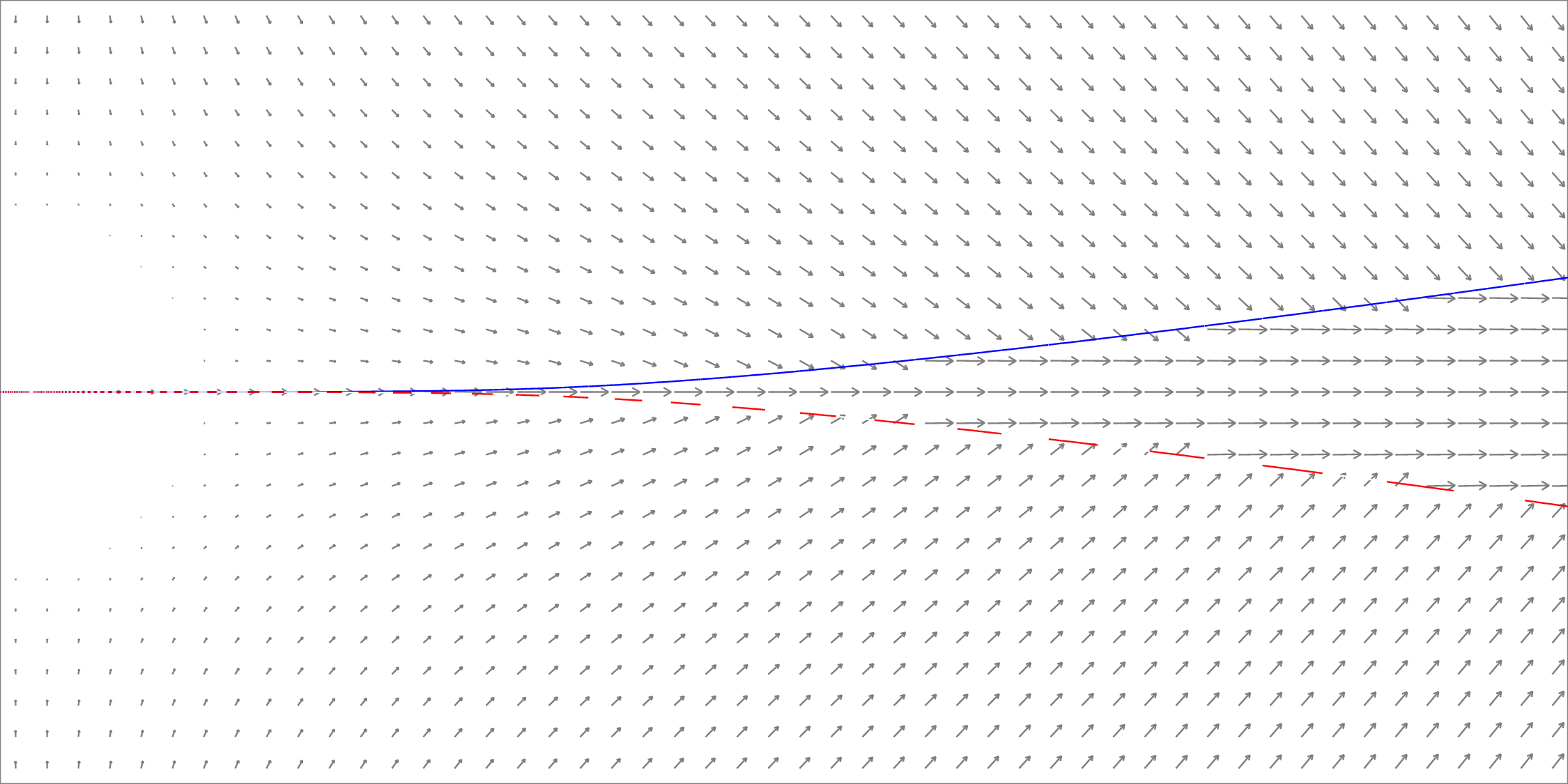}}
  \centerline{\includegraphics[width=.7\linewidth]{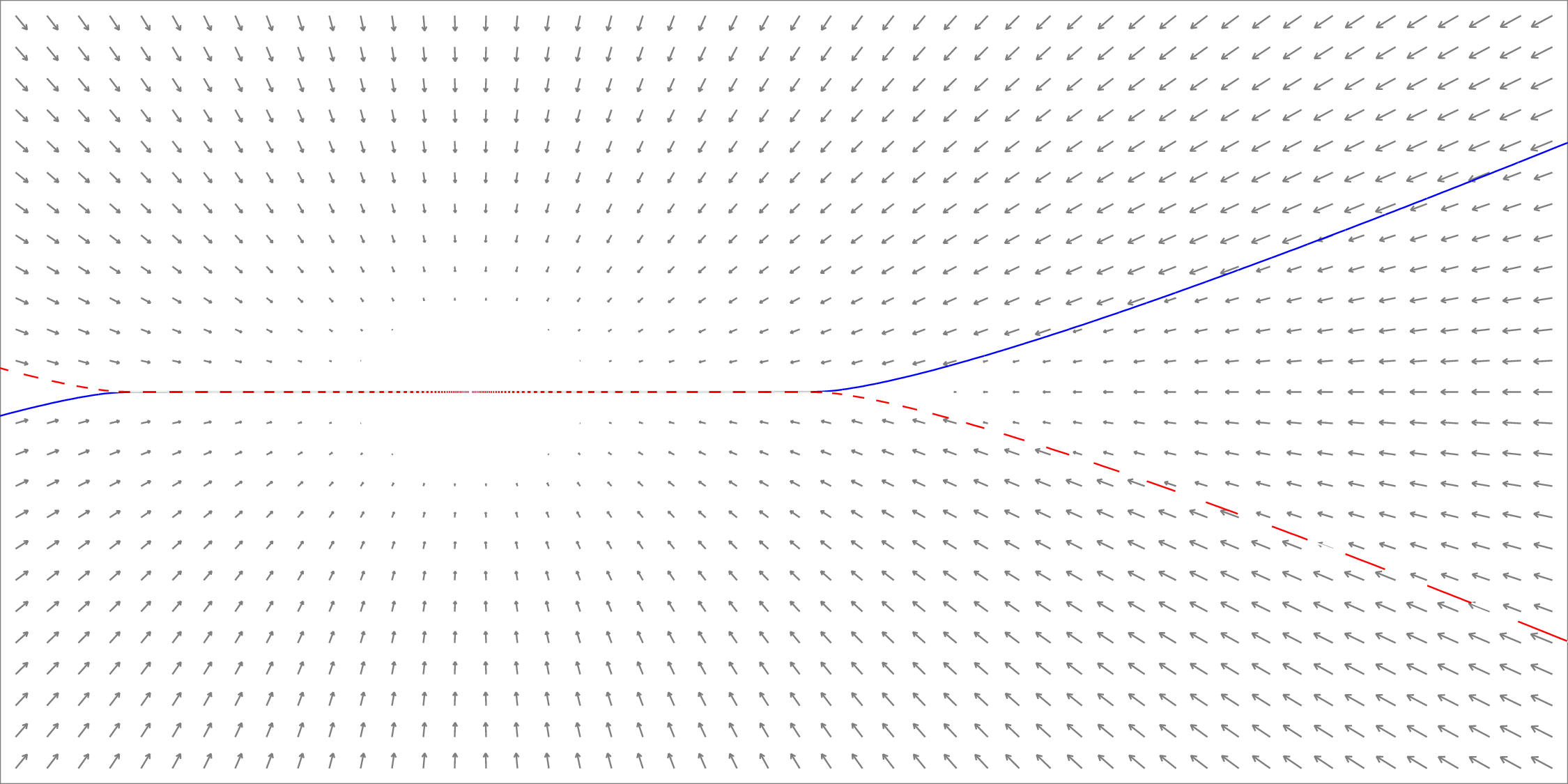}}
  \centerline{\includegraphics[width=.7\linewidth]{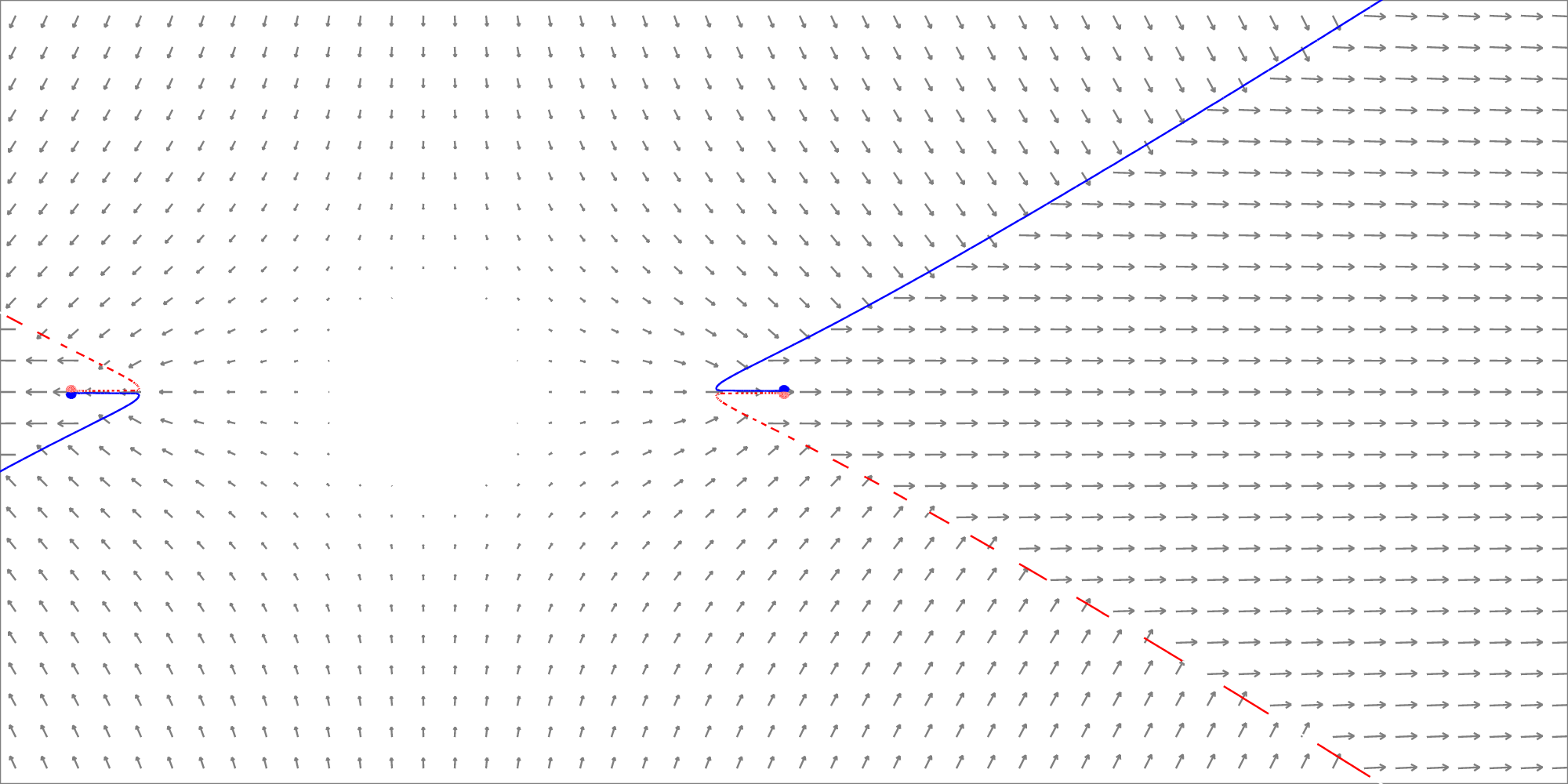}}
  \centerline{\includegraphics[width=.7\linewidth]{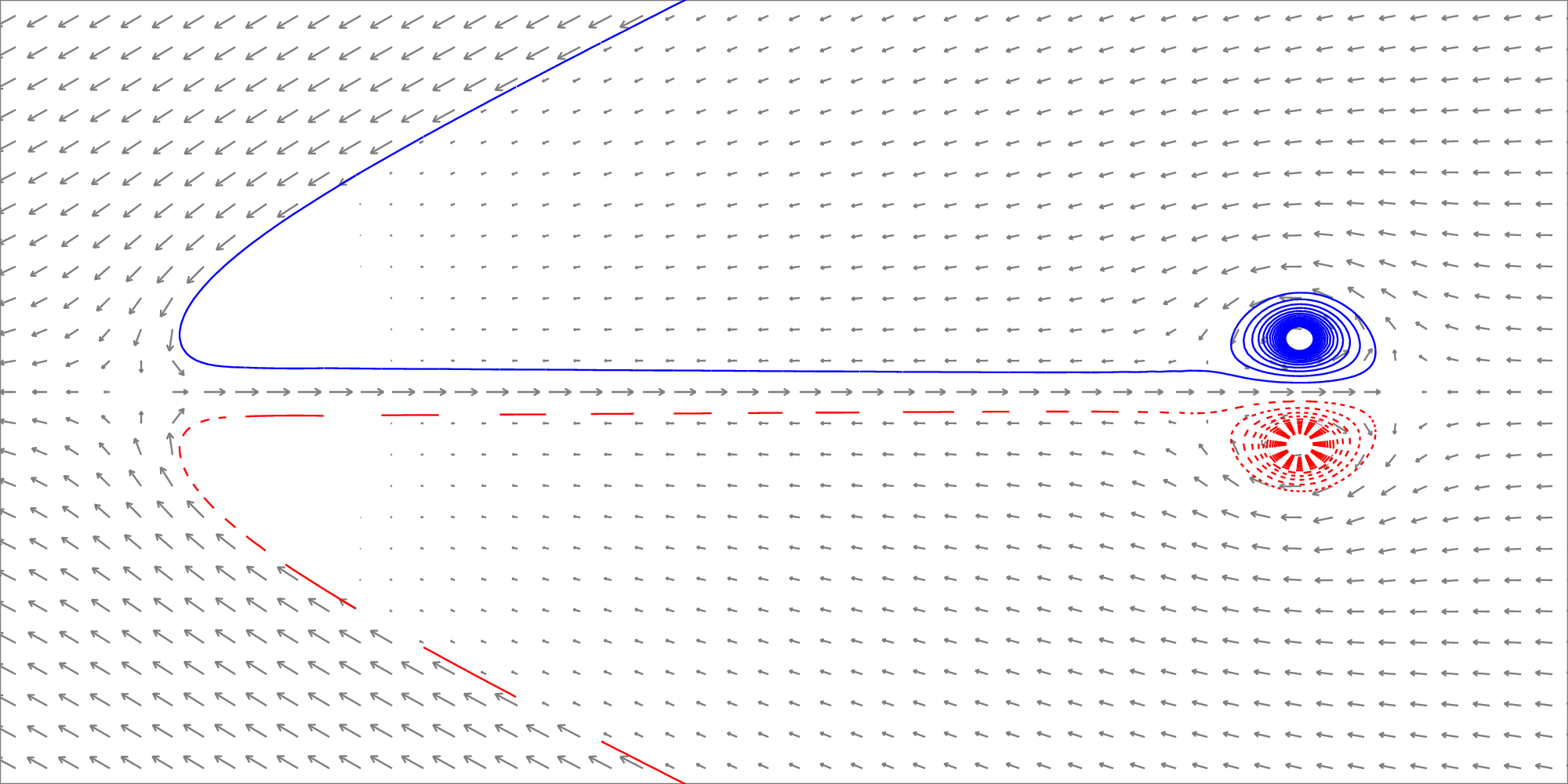}}
  \caption{$\mv=1.3$ with $\Nv=2$ symmetry. Top: for $\phiinf=10^\circ$ a cusp forms (arrows $\cvv$).
    Below top: $\phiinf=25.5^\circ$, near the limit angle (arrows $\cvq$; nearly zero between the sheets where they meet).
    Above bottom: $\phiinf=35^\circ$; instead of a cusp a tiny jet has formed (arrows $\cvv$); bottom: jet detail (arrows $\cvq$; saddle points at jet entrance and exit). }
  \label{fig:cusp}
\end{figure}

Vector fields in all diagrams use a truncated logarithmic scaling to avoid very large or very small arrows. 
Numerical spirals shown in this paper may use an inner cutoff to a central point vortex, 
leaving an empty core in the figures as in fig. \ref{fig:cusp} bottom, physical spirals without cutoff look like fig.\ \ref{fig:spirals} right, with filled-in core.

Our numerical results (see fig.\ \ref{fig:cusp} top) show that vortex cusps exist under some conditions:
if $\mv$ is sufficiently large, and if the data is not too ``large'', which means in the special case of global cusps that $\phiinf$ is sufficiently small, 
below a \defm{limit angle} that depends on $\mv$ and other parameters. 
In cases where cusps do not appear spiral-end jets can be observed instead.

\begin{figure}
  \centerline{\includegraphics[width=.7\linewidth]{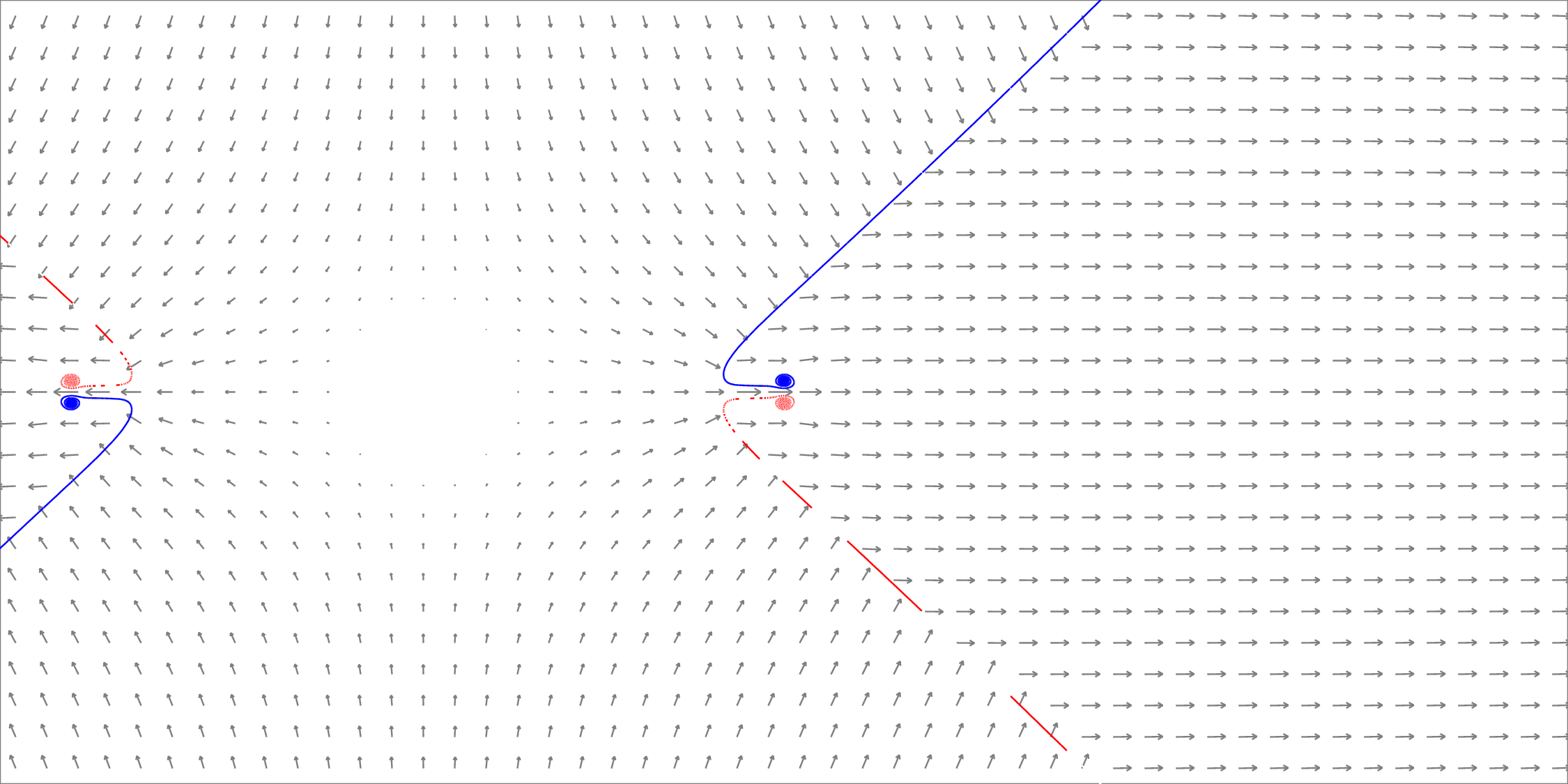}}
  \centerline{\includegraphics[width=.7\linewidth]{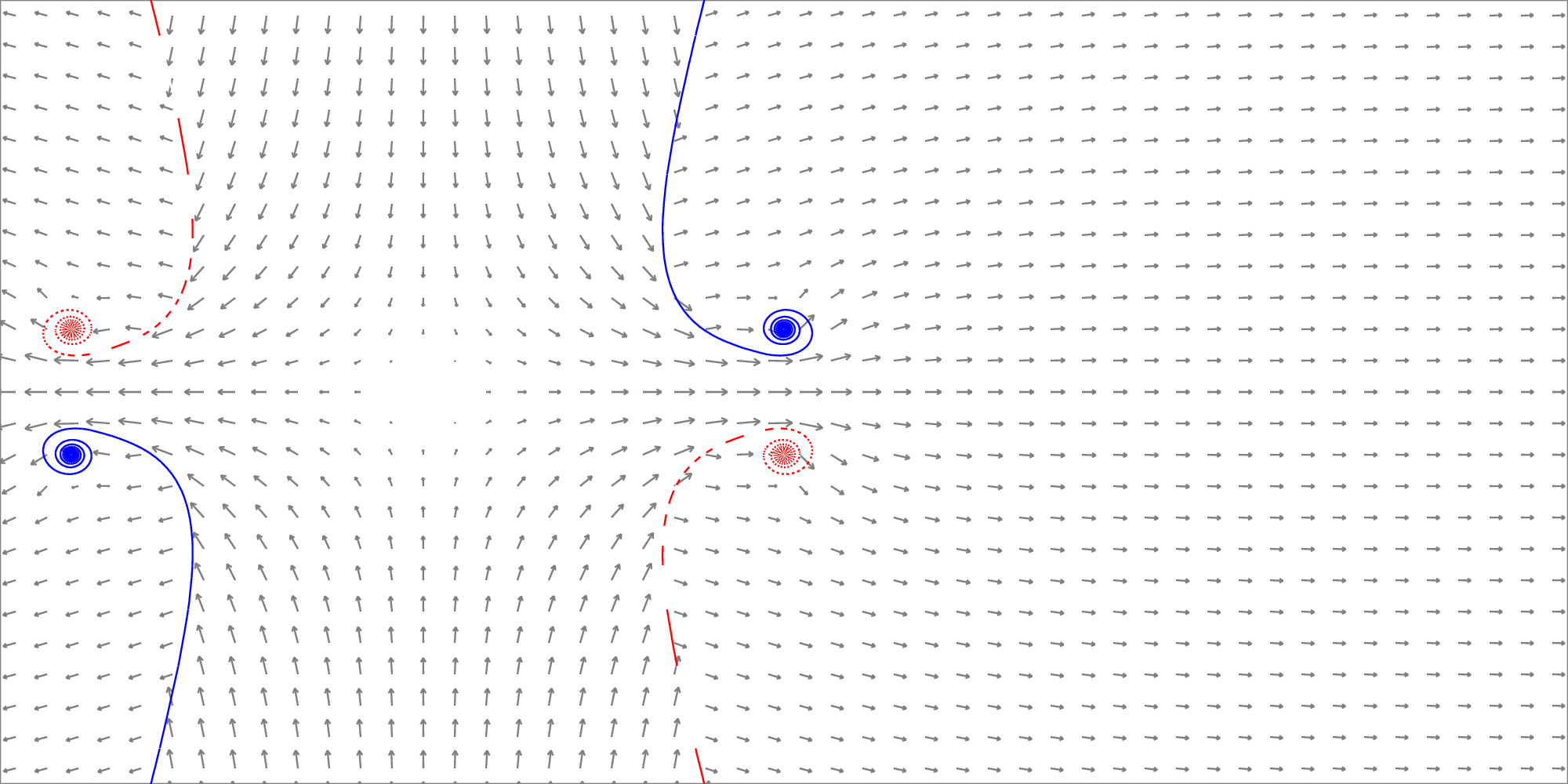}}
  \caption{$\mv=1.3$ with $\Nv=2$ symmetry. Left $\phiinf=48^\circ$, right $\phiinf=77^\circ$, arrows $\cvv$.}
  \label{fig:morejets}
\end{figure}

\subsubsection{Linear and higher degree}

\begin{figure}
  \centerline{\input{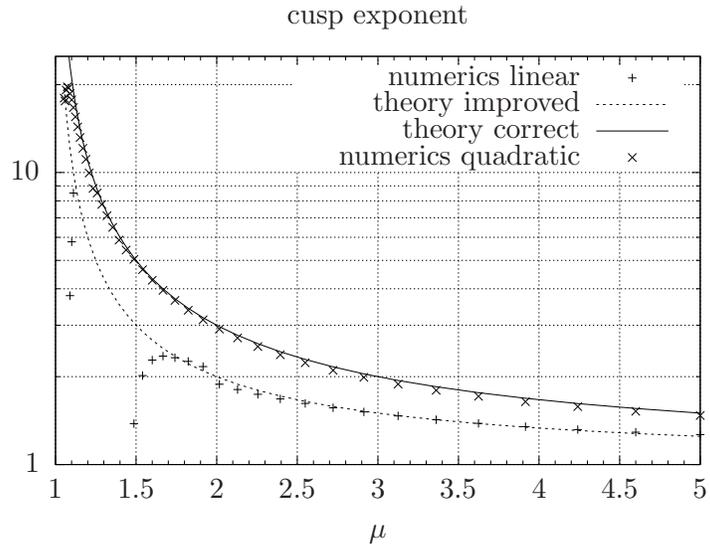}}
  \caption{$\Nv=3$ global cusp exponents from piecewise linear or quadratic numerics}
  \label{fig:N3yex}
\end{figure}

We find that the linear reconstruction is insufficient, qualitatively incorrect and also numerically unstable:
consider the case of global cusps with $\phiinf=1^\circ$ using $\Nv=3$ symmetry, hence $\wec=0$ so that its effects can be suppressed temporarily,
so the ``correct'' cusp exponent $\yex$ is $(\mv+1)/(\mv-1)$ (see \eqref{eq:correct-yex}), the ``improved'' one $\mv/(\mv-1)$ (see \eqref{eq:improved-yex}).
In figure \ref{fig:N3yex} we vary $\mv$ in a wide range from $1$ to $5$ and plot the numerical cusp exponent $\yex$. 

Cusp exponents were estimated by fixing two sufficiently distant points well inside the cusp region of the sheet pair but not too close to the inner cutoff to suppress spurious influences. Since for $\Nv=3$ we know the cusp location is the origin, 
fitting $\cy=C_1\cx^\yex$ (or $\cgam=C_2\cx^\gex$) to the data is after logarithms a simple linear equation for $\log C_i,\yex,\gex$. 

Numerics with linear reconstruction consistently produces numerical solutions with ``improved'' exponent cusps (dotted curve in fig.\ \ref{fig:N3yex}) when $\mv$ is large. 
This is reasonable since the linear approximation is unable to resolve the 2nd derivative $\cgam_{\cx\cx}$ appearing as the extra term in \eqref{eq:correct-yode}
that makes the difference between ``improved'' and ``correct'' exponent. 
In addition it was checked numerically that the sheets produced by the piecewise linear approximation violate conservation of mass $\ndiv\cvq=-2\mv$; 
which is not surprising since conservation of mass arguments also forced us to accept the ``correct'' exponent in \eqref{eq:correct-yex}. 

For $\mv$ below $2$ calculation with linear segments becomes increasingly unstable. 
Numerical instabilities for decreasing $\mv$ have several causes. 
First, cusp exponents $\yex$ increase to infinity; 
then $\cy$ decreases much faster than $\cx$ in the inner cusp so that calculations involving both are prone to large roundoff errors. 
This problem can be reduced by more careful evaluation or higher precision arithmetic. 
Secondly, the limit angle for $\phiinf$ at which cusps change to jets decreases to zero as $\mv$ approaches $1$. 
These two issues are observed for higher degree as well, but thirdly linear approximation also seems inherently more unstable, 
causing more breakdowns in the Newton iteration and requiring smaller time step, 
contrary to the standard expectation that lower-order numerics are more robust. 

Quadratic reconstruction, as well as \emph{higher-degree} ones, are more robust and yield consistent results, 
in particular the same ``correct'' cusp exponent (solid curve in fig.\ \ref{fig:N3yex}, closely matching the ``x'' for quadratic numerics). 
The match is good except near $\mv=1$ where large exponent and cusp-jet transition start to take effect. 

Although we are not concerned with fast numerics here, it is interesting to point out that accurate exponents can be obtained by using quadratic or higher degree 
only for the sheet segments near the evaluation point, but linear segments everywhere else. 
Again this matches our modelling since the cusp exponents and shape are largely governed by an ordinary differential equation from the local part of the BR integral, 
with cancellations largely eliminating non-local contributions. 
This allows considerable numerical acceleration, although for complex flows with additional sheets or other forms of vorticity it may be necessary to detect proximity to them,
changing degree as needed, which then causes differentiability problems affecting Newton iteration and other stability issues. 

In any case we use either quadratic or quadratic-nearby-linear-elsewhere reconstruction for the rest of this paper;
experiments did not show significant differences between them or cubic or higher degrees.

\subsubsection{Limit angle}

\begin{figure}
  \parbox{.47\linewidth}{%
  \centerline{\input{quadcoarse.pstex}}
  \caption{Upper limit on $\phi_\infty$ for existence of $\Nv=3$ global cusps}
  \label{fig:anglemax}%
}\hfil%
  \parbox{.47\linewidth}{%
    \centerline{\includegraphics[width=\linewidth]{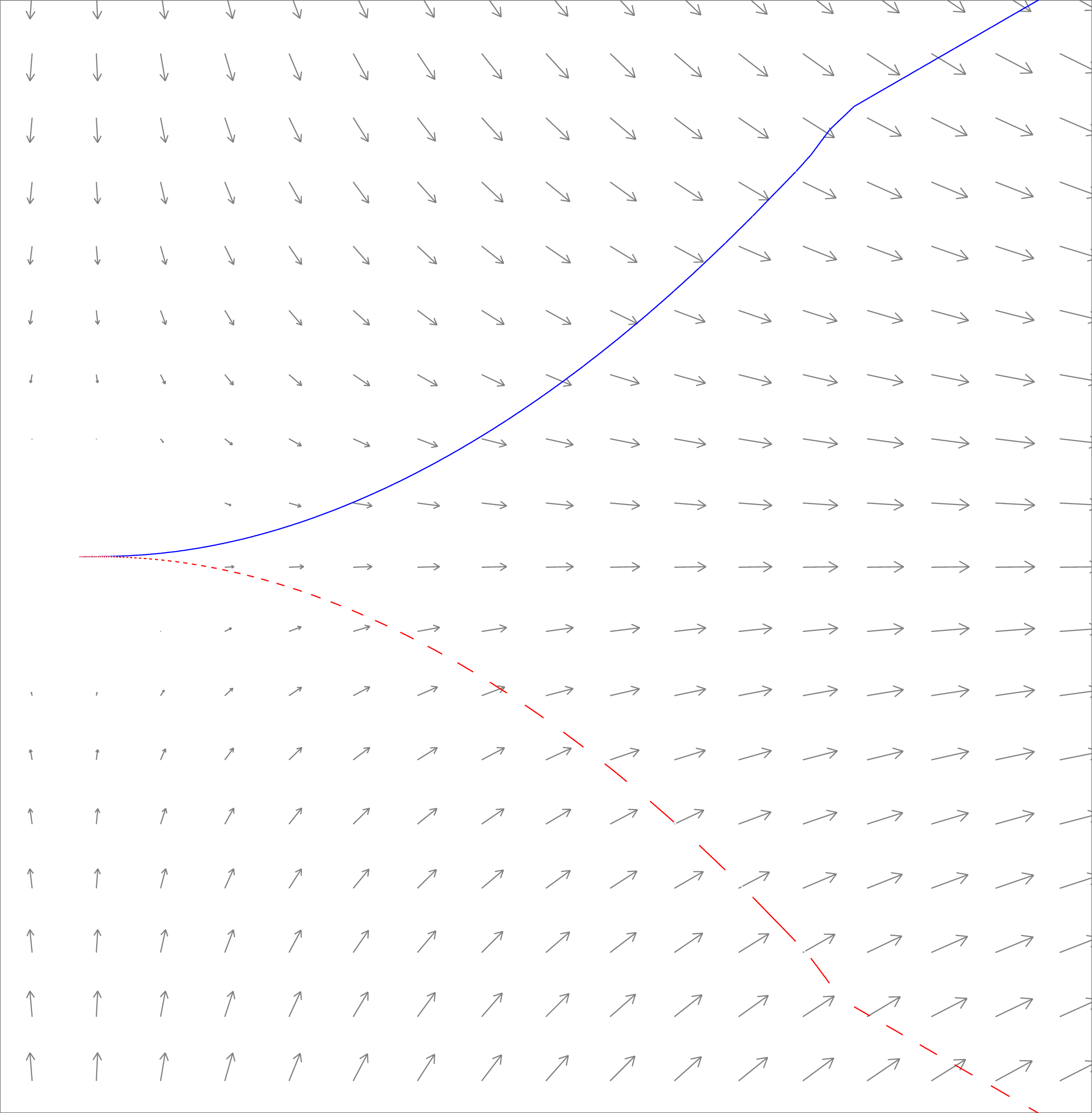}}%
    \caption{%
      $\mv=0.9$, $\Nv=1$ (no symmetry), external $\cwv$ field adjusted to yield $\wec\approx0.3$ (arrows $\vv$). 
    }
    \label{fig:mulessonecusp}%
  }
\end{figure}

The limit angle for $\Nv=3$ global cusps is shown in figure \ref{fig:anglemax}.
Clearly below $\mv=1.3$ the limit angles become rather restrictive, converging to $0$ as $\mv\dnconv 1$. 
For $\mv\leq 1.05$, near the left side, large cusp exponents cause unreliable calculations.
The limit angle diagram for $\Nv=2$ and other cases is similar; 
in those cases $\wec\conv0$ along with $\mv\dnconv 1$ so that it is \emph{not} possible to obtain cusps at or below $\mv=1$. 

However, it is possible to see \emph{local} cusps for $\mv\leq 1$ if a large strain coefficient $\wec$ is generated by other vorticity in the similarity plane
or by externally imposed harmonic velocity fields. In figure \ref{fig:mulessonecusp} we add to the BR integral velocity an additional harmonic velocity field to obtain $\wec\approx0.3$. 
The two straight segments on the right of fig.\ \ref{fig:mulessonecusp} are fixed, only the curved parts solve the BR equations. 
(The pronounced corner between straight and curved segments was inconsequential.) It is possible to achieve cusps at $\mu=0.9$ and lower, 
at fairly ``large'' angles between the straight segments. 

The problems of linear reconstruction are also important for theoretical reasons. 
It is natural to look for initial data that permits more than one self-similar solution, for example two jets or jets and cusps coexisting in a ``hysteresis'' region
of the parameter space. Sufficiently inaccurate numerical methods do suggest that jets may exist below the limit angle at which cusps disappear.

\subsubsection{$\cy$,$\cx$ exponents with strain}
\label{section:xyexponent}

\begin{figure}
  \centerline{\input{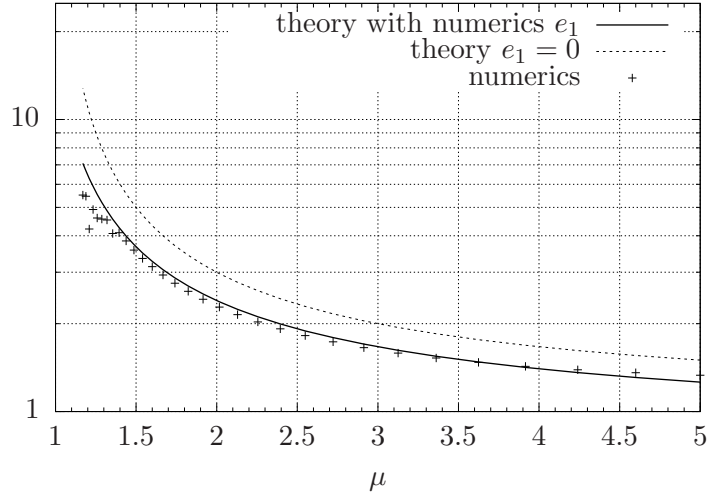}}
  \caption{Numerical cusp exponents $\yex$ in presence of saddle strain}
  \label{fig:wecyex}
\end{figure}

We also wish to demonstrate the effect of the ``saddle coefficient'' $\wec$ on cusp exponents.
Calculations using $\Nv=1$, i.e.\ no symmetry, cause the velocity integral to diverge if $\mv\geq 1$, which is a dilemma since we do not generally expect cusps for most $\mv<1$. 
The divergence could be controlled by numerically subtracting the velocity induced at $z=0$ from every other velocity, 
i.e.\ adjusting the velocity integral by a large constant that becomes infinity as the outer cutoff is increased to infinity. 
But the results did not differ much from the alternative, using $\Nv=2$, i.e.\ eliminating the divergence by symmetry, 
which is the only case we present here (fig.\ \ref{fig:wecyex}).

We used a cusp with $\phi_\infty=10^\circ$; since $\wec$ is mostly generated by the outer non-cusp part of the sheet, a smaller angle would cause a relatively
small and harder-to-estimate $\wec$, while a larger angle causes cusps to cease to exist for $\mv$ far above $1$ 
(which is one of the reasons for the inaccurate results seen near $\mv=1$ in the diagram, the other reason being that the larger the cusp exponent the more
their numerical estimation is affected by numerical roundoff errors from tiny $\cy$).
The coefficient $\wec$ was estimated numerically by evaluating the velocity integral derivative in some point near the cusp but \emph{outside} the sheet pairs 
where their cusp parts undergo strong cancellation and \emph{most} of the velocity is induced by far-away sheet parts. 

We evaluate the ``correct'' cusp exponent $\yex=(\mv+1-\wec)/(\mv-1+\wec)$ from \eqref{eq:correct-yex} 
once with $\wec=0$ (dotted curve in fig.\ \ref{fig:wecyex}) and once with the numerically estimated $\wec$ (solid curve); 
none of these $\wec$ exceeded $\half$. 
The numerical measurements fit the solid curve well, so we conclude that our modelling likely yielded the correct $\wec$ dependence of the cusp exponent. 
For calculating the diagram we used an unncessarily coarse discretization and smallish outer cutoff to display the errors more clearly. 
The numerical cusp exponents fit the solid curve closely, except near the $\mv\dnconv1$ end where large cusp exponents and transition to jetting cause now-familiar inaccuracy; 
the fit at large $\mv$ can be improved further by taking larger outer cutoffs. 
This corresponds to $\wec$ being more sensitive to outer parts of the sheets when $\mv$ is large. 
To improve the fit from $\mv=1.5$ to $\mv=3$ it was sufficient to refine the discretization alone.

\subsubsection{$\cgam$,$\cx$ exponents with strain}
\label{section:gxexponent}

Finally, in figure \ref{fig:xgexponent} we consider the exponent $\gex$ in the $\cgam=C\cxd^\gex+..$ relationship in the cusp limit. 
Formula \eqref{eq:gex} distinguishes the cases $\wec<\half$ and $\wec>\half$. 
The strain coefficient $\wec$ induced by the non-cusp parts of the sheet increases monotonically 
as the infinity angle $2\phiinf$ between the outer parts increases;
$\wec$ \emph{does} exceed $\half$ if $\mv$ is chosen sufficiently large (with $\mv=2.5$ chosen for computing fig.\ \ref{fig:xgexponent}), 
allowing not only a larger angle but also using that for $\mv\upconv\infty$
the effects of the outer sheet in the velocity integral become ever stronger. 

\begin{figure}
  \centerline{\input{xgexponent.pstex}}
  \caption{Solid curve: theoretical $\gex$ \eqref{eq:gx-nonstandard};
    dotted curves: numerically estimated $(\wec,\gex)$.}
  \label{fig:xgexponent}
\end{figure}

The results, again using a numerically estimated $\wec$, show a very close match between formula \eqref{eq:gex} (solid curve) 
and the numerically determined $\cgam,\cx$
exponent $\gex$ (dashed and dotted curves). Near $\wec=\half$ the results are almost $0.02$ apart. Errors decrease rapidly at a distance from $\wec=\half$ as the 
inner cutoff is taken smaller and simultaneously the initial $-d\cxd/\cxd$ stepsize is reduced
(the dotted curve has inner cutoff improved by a factor $1/10$ and resolution by $1/2$). 
But convergence is very slow near $\wec\approx\half$, which is natural: 
not only is it difficult to distinguish a power law exponent $\gex=0.5$ from $\gex=0.49$ numerically, 
but as $\wec\upconv\half$ the \emph{coefficient} $C$ in $\cgam=C\cx^{0.5}+...$ \emph{also} vanishes (see \eqref{eq:gxconst}), 
meaning the leading term is dominated by the next one for all but astronomically small inner cutoffs. 

We conclude that our formulas \eqref{eq:gex} and \eqref{eq:correct-yex} for $\cgam,\cx$ and $\cy,\cx$ exponents are probably the physically correct exponents,
with close match between numerical data and heuristic derivations. We emphasize again that the $\cy,\cx$ exponent is completely determined by the conservation of mass argument,
so that the point of attack for criticism is confined to the $\cgam,\cx$ exponent and the ODE model and simplifying assumptions that led to it.

\subsubsection{Clockwise upper circulation}
\label{section:clockwise}

\begin{figure}
  \centerline{\includegraphics[width=.47\linewidth]{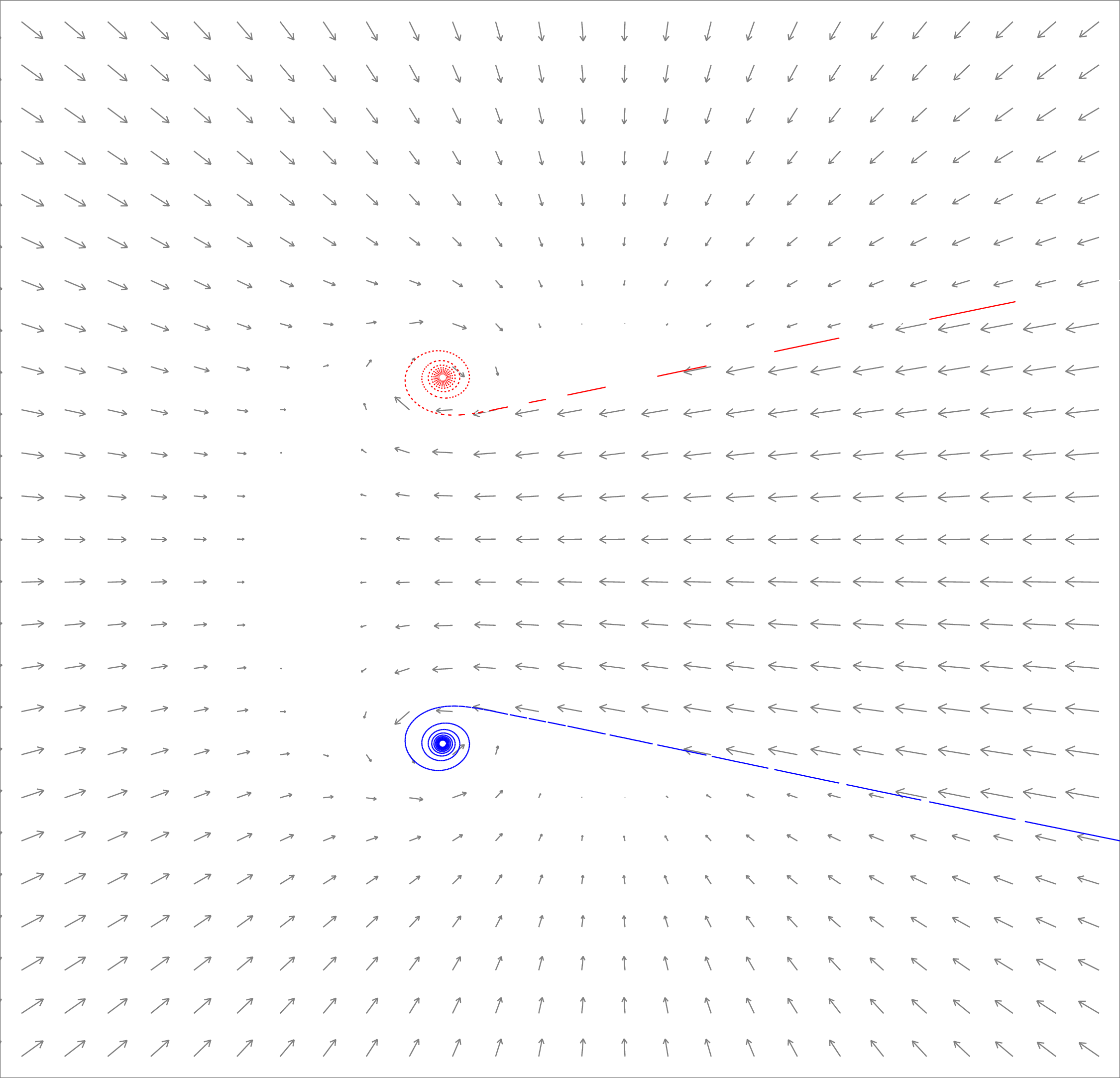}}%
    \caption{Jets tend to form for upper sheets with clockwise circulation ($\mv=1.3$, $\Nv=1$ (unsymmetric), $\phiinf=10^\circ$, arrows $\cvq$). }
    \label{fig:clockwise-jet}%
\end{figure}

\begin{figure}
  \centerline{\includegraphics[width=.7\linewidth]{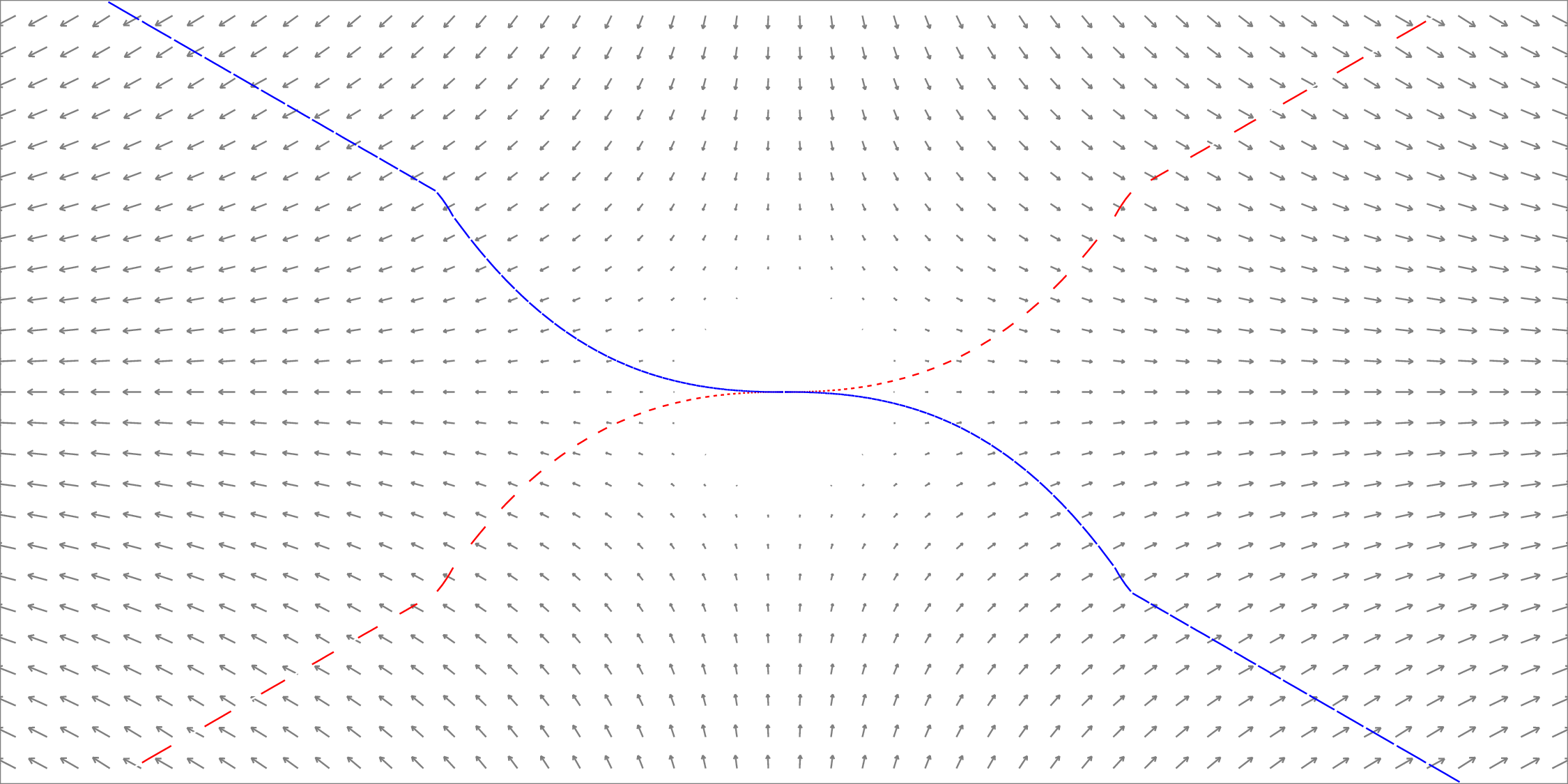}}%
  \caption{A local cusp with reversed circulation (clockwise on upper right sheet), made possible by strain coefficient $\wec>\half$; arrows $\cvv$.}%
  \label{fig:negcirc}%
\end{figure}

For $\wec<\half$ our $\cx$ ODE discussion in section \ref{section:cx-solve} yielded solutions $\cgam_\cxd = (\half-\wec)\cxd + ... $; note $\half-\wec>0$ so that $\cgam_\cxd>0$ near the cusp, corresponding to counterclockwise circulation on the upper sheet near cusps opening to the right. 
For clockwise circulation jets rather than cusps will be observed, as in the local flow in figure \ref{fig:clockwise-jet} ($\mv=1.3$, $\Nv=1$). 

However, if $\wec>\half$, then section \ref{section:cx-solve} permitted solutions $\cgam_\cxd = C \cxd^{\gex-1} + ... $ for $\gex-1>1$ (see \eqref{eq:gex}), with \emph{arbitrary} $C$. 
In particular we may choose $C<0$, allowing clockwise circulation on the upper sheet at the cusp. 
Numerically it can be confirmed that such solutions are possible: figure \ref{fig:negcirc} shows a local $\Nv=2$ symmetric cusp with $\mv=1.5$ and $\wec$ adjusted to $0.7$ by adding an external harmonic velocity field $w=cz$
with suitable constant $c$. Again the straight outer segments are fixed chosen data, whereas the curved inner parts are numerical Birkhoff-Rott solutions forming a cusp in the origin. 

Generally the case of global flows with clockwise upper sheet circulation and infinity angle $\phiinf$ is equivalent to counterclockwise circulation and infinity angle $360^\circ/\Nv-\phiinf$.
For example the jets in fig.\ \ref{fig:morejets} with $\phiinf=48^\circ$ and $\phiinf=77^\circ$ can be rotated $90^\circ$ to yield solutions for the clockwise upper sheet circulation case for $\phiinf=42^\circ$ and $\phiinf=13^\circ$.

\subsubsection{Jetting and near-limit-angles}
\label{section:jetting}

As $\phiinf$ approaches the limit angle from below the cusp attains the $Y$ shape in fig. \ref{fig:cusp} second from top. Between the sheets near the meeting point of the Y 
the pseudo-velocity $\cvq$ approaches zero. We were not able to discerm by numerical experiments whether it reaches zero or whether 
transition to jets occurs earlier. This would be of interest for obtaining formulas or criteria predicting the limit angle. 

A natural idea is to consider where the cusp fails to exist because the denominator $\mv-1+\wec$ of the cusp exponent $\yex$ in \eqref{eq:correct-yex} crosses zero. 
But this ``local'' criterion is easily refuted once we consider $\Nv=3$ 
where $\wec=0$ due to cancellation so that cusp-jet transition would always occur at $\mv=1$, clearly at odds with our numerical observations (see fig. \ref{fig:anglemax}).

Hence cusp-jet transition is a more non-local effect, influenced strongly by the ambient flow especially near the meeting point rather than the cusp. 
Better criteria may be based on conservation of mass arguments, 
possibly using $\qxi$ or similar variables approaching zero. 
It is not clear whether we should expect a single criterion to be suitable for large classes of flows. 

Above the limit angle the cusp ceases to exist; sheet and mirror image separate with ends flipping rightward in between them and forming a spiral-ends jet;
see fig.\ \ref{fig:cusp} bottom two as well as fig.\ \ref{fig:morejets}. 
As $\phiinf$ is increased this flow pattern persists, with sheets separating further.
On the other hand as $\phiinf$ converges to the limit angle from above the distance of the two sheets converges to zero from above. 

For $\mv=1.3$ the limit angle is near $\phiinf=25.5^\circ$, but already at $\phiinf=48^\circ$ (fig.\ \ref{fig:morejets} top), 
and even more so at $\phiinf=35^\circ$ (fig.\ \ref{fig:cusp} bottom two) which is still almost $10^\circ$ above the limit angle, the jets are very small. 
Indeed numerical calculations show that distance between the spiral centers decreases \emph{quadratically} in the distance of $\phiinf$ from limit angle. 
Due to the instability of vortex sheets in general and the smallness of the jet in particular, 
computing jet solutions near the limit angle is even more challenging than computing vortex cusps.

\subsubsection{Relation to Mach reflection}

\cite{henderson-vasilev-bendor-elperin-2003} and \cite{vasilev-bendor-elperin-henderson-2004} 
have begun a theoretical and numerical study on jetting in Mach reflection of shock waves for compressible flow, 
whose quasi-incompressible regions correspond to $\mv=1$ in our context. 
It is well-established (see \cite{henderson-menikoff}, \cite{serre-hbfluidmech} and \cite{elling-tripleshock}) that the vortex sheets generated by single Mach reflection 
(fig.\ \ref{fig:mr} right) have counterclockwise circulation. 
This is very important because it allows, by our analysis above, the presence of cusps even if the ambient flow generates only
weak or negative-sign strain coefficients $\wec$. 

By regarding sheet-wall interaction as an essentially incompressible phenomenon 
and adapting vortex sheet numerics to obtain better resolution, 
we found that when jets occur they may be rather small, possibly small enough to be obscured by numerical or physical dissipation. 
Besides, for $\mv=1$ we find that cusps are possible only if the ambient flow generates a positive (but not too large) strain coefficient $\wec$. 
Even then it is necessary that the outer (non-cusp) part of the sheet forms a rather small angle $\phiinf$ to the wall, increasingly smaller as $\wec$ approaches zero from above.
This suggests that many flows that appear to be cusps are really jets in disguise. 
For instance $\Nv$-symmetric Mach reflections with $\Nv\geq 3$ must have $\wec=0$ for cusps in the origin which is not possible if $\mv=1$. 
The problem is compounded by the rather large cusp exponents that result from near-zeros of the denominator $\mv-1+\wec$ in \eqref{eq:correct-yex} 
when $\mv=1$ and $\wec>0$ is small; such exponents cause most of the cusp part to be very close to the wall, 
making it appear at large scale that the sheet meets the wall at a positive angle. 

Shock capturing methods on non-adapted grids are much less accurate than the vortex methods employed here. 
In calculations small jets are easily obscured by numerical or physical boundary layers and by general inaccuracy. 
Of course in an exactly self-similar flow the inviscid features of the flow often outgrow the viscous ones;
whether the jet-cusp distinction is important would depend on the intrinsic length and time scales of the application at hand. 
Small-scale cusp-jet differences may also make themselves felt in attachment vs.\ separation of boundary layers that cause differences at a much larger scale.

A natural and relatively cheap prediction for cusp vs.\ jet is to measure in the output of shock capturing calculations
the sheet-wall angle $\phiinf$ and the velocity $\wve$ outside the hypothetical cusp;
this requires sufficiently high resolution for a significant region relatively free of boundary effects and with Mach number low enough that flow can be treated as incompressible. 
A negative $\wec$ immediately rules out cusps; given a positive value either the model ODE or 
(for larger $\phiinf$ where our approximations are inaccurate outside the inner cusp region) 
numerical solution of the full Birkhoff-Rott equation can determine whether jetting occurs. 
This approach avoids modification of existing numerical software which is more time-consuming even when the source code is available.

\section{Additional considerations}

\subsection{$e$ asymptotics}
\label{section:e-depth}%

During modelling starting with \eqref{eq:upperpoint} we have neglected the velocities induced in points $\cz$ by ``non-near'' other points $\cz'$ 
when both are located in the cusp region of the sheets. 
Of course despite strong cancellation these contributions are also nonzero; here we perform a crude check of the decay rate.

Assume for simplicity the cusp $\cxu$ has been shifted to $0$. 
We regard $\cz'$ as ``not near'' $\cz$ if $|\cx'-\cx|\geq\delta\cx$ for some constant $\delta>0$, with $0<\cx,\cx'<\overline\cx$ where $\overline\cx$ marks the ``boundary'' of the cusp region, and consider the $\cx,\cx'\conv\cxu=0$ limit. First consider the $0<\cx'<(1-\delta)\cx$ part of the Birkhoff-Rott integral:
\[ w(\cz) 
= \int_0^{(1-\delta)\cx} (\frac1{\cz-\cz'} - \frac1{\cz-{\cz'}^*}) \frac{\cgam_\cx(\cx')d\cx'}{2\pi i} 
\eqs \int_0^{(1-\delta)\cx}  \frac{\cy'\cgam_\cx(\cx')d\cx'}{(\cx-\cx'+i\cy)^2+(\cy')^2} 
\] 
We obtained solutions of type $\cgam\eqs\cx^\gex$ and $\cy\eqs \cx^\yex$ with $\yex>1$, so $\cy,\cy'$ in the denominator are dominated by $\cx,\cx'$ (if non-near):
\[ w
 \eqs \int_0^{(1-\delta)\cx} \frac{{\cx'}^\yex {\cx'}^{\gex-1}  d\cx'}{(\cx-\cx')^2} .
\] 
Non-near means $(\cx-\cx')^{2}\sim(\cx')^{2}$, so
\[ 
w \eqs \int_0^{(1-\delta)\cx} \frac{{\cx'}^\yex}{\cx'^2} {\cx'}^{\gex-1} d\cx' \eqs \cx^{\yex+\gex-2}.
\] 
assuming the last exponent of $\cx$ is positive. 
Our modelling results yielded $2\leq\gex<\infty$; proper cusps have $1<\yex<\infty$, 
so the exponent is in fact $>1$. 

However, if $\gex=2$ and $\yex=1+\eps$, then the exponent is $1+\eps$, just barely above $1$. 
Such $\gex,\yex$ do occur; consider for example $\Nv\geq 3$ symmetry so that $\wec=0$ and thus $\gex=2$, then take $\mv\conv\infty$ so that 
exponent $\yex+\gex-2=(\mv+1-\wec)/(\mv-1+\wec)\dnconv 1$; already for $\mv>3$ the exponent is below $2$. 

For the integral over $(1+\delta)\cx<\cx'<\overline\cx$ the only difference is a term from the $\overline\cx$ boundary that is constant in $\cx$, 
hence can be regarded as part of $\wez$.

We conclude that our approximation $\wve(\cz)=\wez+\wec cz+...$ was reasonable, but that expanding to 
\[ \wve(\cz) = \wez + \wec\cz + \wve_2\cz^2 + ... \] 
or deeper is not always reasonable since the quadratic term may be dominated by effects we already neglected.

\subsection{$\cgam_*>0$ cusps}
\label{section:xcusps}

In section \ref{section:constraints} we deferred the case $\cgam_*>0$; we revisit it here, showing that such cusps may occur but are ``non-generic''. In $t,\xx$ coordinates $\tgam$ is merely a parameter that can be changed by an additive constant without consequences. But the $t$-scaled $\cgam$ in in similarity coordinates is not as arbitrary since it corresponds the rate at which circulation passes through a point 
through self-similar movement alone.
Recall \eqref{eq:cgse} that $\cgam_*>0$ means $\mv\cxu\neq\wez$. 
Take $\pd\cxd$ of \eqref{eq:qxig}: again we use 
\begin{alignat}{5}
  \cgam_\cxd = \qxi + \mv\cx - \wez - \wec\cxd  \label{eq:gxtwo}
\end{alignat}
to calculate
\begin{alignat}{5}
  0 
  &= \qxi\qxi_\cxd + (\mv\cx-\wez-\wec\cxd)(\wec-\mv) + (2\mv-1)\cgam_\cxd
  \notag\\&= \qxi\qxi_\cxd + (\mv\cx-\wez-\wec\cxd)(\wec+\mv-1) + (2\mv-1)\qxi
  \notag\\&\overset{\cx=\cxu+\cxd}{=} (\frac{(\qxi)^2}{2})_\cxd + (2\mv-1)\qxi + (\mv-\wec)\cxd(\wec+\mv-1) + \subeq{(\mv\cxu-\wez)}{\neq0}(\wec+\mv-1)
  \label{eq:qq}
\end{alignat}
The middle terms decay as $\cxd\dnconv 0$, but the last term is nonzero unless the scalar constraint
\begin{alignat}{5}
  \wec &= 1-\mv  \label{eq:econs}
\end{alignat}
is satisfied. Assume it is not. Then \eqref{eq:qq} yields
\begin{alignat}{5}
  (\frac{(\qxi)^2}{2})_\cxd &= C + o(1) \quad\text{as $\cxd\dnconv 0$} \label{eq:qqq}
\end{alignat}
for a nonzero constant $C$ (necessarily positive as evident below). Integrate:
\begin{alignat}{5}
  \qxi &= -\sqrt{2C\cxd} + o(\cxd^{1/2}) , \label{eq:qxiasy}
\end{alignat}
where $\qxi\conv0$ as $\cxd\dnconv 0$ eliminated the integration constant; $-$ is chosen since $\qxi<0$ (derived from conservation of mass). Then \eqref{eq:qqq} shows
\begin{alignat}{5}
  \qxi_\cxd = -\sqrt{\frac{C}{2\cxd}} + o(\cxd^{-1/2}). \label{eq:qxisqrt}
\end{alignat}
Now consider our ``correct'' model \eqref{eq:correct-yode-qxi} for the upper sheet $\cy=\cy(\cx)$:
\begin{alignat*}{5} 
  \frac{\cy_\cxd}{\cy}
  &=
  \frac{ 2\mv+\qxi_\cxd }{ -\qxi } 
\end{alignat*}
By \eqref{eq:qxisqrt} the numerator is dominated by $\qxi_\cxd$; using also \eqref{eq:qxiasy} we find
\begin{alignat*}{5}
  (\log\cy)_\cxd
  &=
  - \frac1{2\cxd} + o(\cxd^{-1}) 
  \qiq
  \cy \sim \frac1{\sqrt\cxd} 
\end{alignat*}
Hence a cusp cannot form; $\cy$ necessarily blows up as the cusp is approached. 

We emphasize that this blowup is relatively mild and easily overlooked in low-accuracy numerical calculations
where it may appear falsely that a cusp might form upon further refinement. 
Besides, blowup is entirely due to the $\cgam_{\cxd\cxd}$ term (corresponding to $\qxi_\cxd$ above) which is only present in the ``correct'' model,
but not in the other two; numerics with less than quadratic approximation would not recognize blowup, producing false cusps! 

On the other hand \emph{if} the constraint \eqref{eq:econs} is satisfied, there may be a chance for cusps to form. 
A candidate for such cusps are the flows arising as the maximal-$\phiinf$ limit, see fig.\ \ref{fig:cusp} right: 
in that limit the ``standard'' $\cgam_*=0$ cusp collapses to two coinciding horizontal lines; 
the region where they separate appears to be a cusp as well, but necessarily with $\cgam_*>0$ as observed in numerics. 
However, further modelling requires deeper Taylor expansions of $\wve$ which are not always justified, as explained in the previous section. 
Besides, 
the theoretical condition \eqref{eq:econs} imposes a scalar constraint on the parameters;
similarly $\phiinf$ being the limit angle is a ``borderline'' case, not robust under parameter changes.
Since $\cgam_*>0$ cusps do not obviously appear in other circumstances our investigation appears to have reached the point of diminishing returns.

\providecommand{\bysame}{\leavevmode\hbox to3em{\hrulefill}\thinspace}
\providecommand{\MR}{\relax\ifhmode\unskip\space\fi MR }
\providecommand{\MRhref}[2]{%
  \href{http://www.ams.org/mathscinet-getitem?mr=#1}{#2}
}
\providecommand{\href}[2]{#2}

\end{document}